\documentclass[12pt]{iopart}
\usepackage{iopams, graphicx, color, mathrsfs, bbm, ulem}
\usepackage{epstopdf}
\usepackage[pdftitle={Phase space entanglement}]{hyperref}

\graphicspath{{./}{./figures/}}
\newcommand{\beq}{\begin{equation}}
\newcommand{\eeq}{\end{equation}}

\newcommand{\eq}[1]{Eq.~(\ref{#1})}

\newcommand{\rank}{\text{rank}\,}

\newcommand{\abs}[1]{\left| #1 \right|}

\newcommand{\inner}[2]{\left\langle #1, #2 \right\rangle}

\newcommand{\vecenv}[2]{\left( \begin{array}{c} #1 \\ #2 \end{array} \right)}
\newcommand{\diag}[1]{\text{diag}\left\{ #1 \right\}}
\newcommand{\spec}[1]{\sigma \left[ #1 \right]}

\newcommand{\wt}[1]{\widetilde{#1}}

\newcommand{\cmplx}{\mathbb{C}}
\newcommand{\intg}{\mathbb{Z}}
\newcommand{\real}{\mathbb{R}}

\newcommand{\uvec}{\mathbf{e}}
\newcommand{\sech}{\text{sech}}

\newcommand{\id}{\mathbbm{1}}
\newcommand{\nullv}{\mathbf{0}}

\newcommand{\proj}{\mathcal{P}}

\newcommand{\uop}{\mathcal{U}}

\newcommand{\U}{\mathrm{U}}

\newcommand{\SO}{\mathrm{SO}}
\newcommand{\Sp}{\mathrm{Sp}}
\newcommand{\GL}{\mathrm{GL}}

\newcommand{\ISp}{\mathrm{ISp}}

\newcommand{\hlt}{\mathcal{H}}

\newcommand{\ve}{\varepsilon}

\newcommand{\bra}[1]{\langle #1 |}
\newcommand{\ket}[1]{| #1 \rangle}
\newcommand{\braket}[2]{\langle #1 | #2 \rangle}

\renewcommand{\a}{a}
\newcommand{\ad}{a^\dagger}

\newcommand{\vA}{\mathbf{A}}

\newcommand{\vS}{\mathbf{S}}

\newcommand{\va}{\mathbf{a}}
\newcommand{\vb}{\mathbf{b}}

\newcommand{\vm}{\mathbf{m}}

\newcommand{\vp}{\mathbf{p}}

\newcommand{\vu}{\mathbf{u}}
\newcommand{\vv}{\mathbf{v}}

\newcommand{\vx}{\mathbf{x}}

\newcommand{\bal}{{\boldsymbol{\alpha}}}

\newcommand{\bga}{{\boldsymbol{\gamma}}}

\newcommand{\bom}{\boldsymbol{\omega}}

\newcommand{\hilbert}{\mathscr{H}}

\newcommand{\cn}{\colon}
\newcommand{\dg}{{\dagger}}
\newcommand{\pdg}{{\phantom{\dagger}}}
\newcommand{\viz}{\emph{viz}}

\newcommand{\hypgeo}{{\phantom{.}_2F_1}}

\newcommand{\ph}[1]{{ \color{red} [??]}}

\newcommand{\text}[1]{{\rm #1}}
\newcommand{\binom}[2]{\phantom{.}^{#1}C_{#2}}
\newcommand{\implies}{\Longrightarrow}

\newcommand{\overlap}{\mathcal{O}}
\newcommand{\overlapHO}{\mathfrak{O}}
\newcommand{\eig}{\mu}
\newcommand{\wige}{\hat{\mathfrak{e}}}
\newcommand{\wigE}{\hat{\mathfrak{E}}}

\newcommand{\sA}{\text{A}}
\newcommand{\sB}{\text{B}}
\newcommand{\sE}{\text{E}}
\newcommand{\coord}{z}
\renewcommand{\emph}[1]{{\it #1}}
\newcommand{\rmo}{\mathrm{o}}
\newcommand{\fA}{\mathfrak{A}}
\newcommand{\fb}{\mathfrak{b}}

\renewcommand{\leq}{\leqslant}
\renewcommand{\geq}{\geqslant}

\begin{document}
\title{Phase Space Entanglement Spectrum}

\author{Vatsal Dwivedi$^1$, Victor Chua$^2$}
\address{$^1$Institute for Theoretical Physics, University of Cologne, \\ Z\"ulpicher Stra\ss e 77a, 50937 Cologne, Germany}
\address{$^2$Department of Physics, University of Basel, \\ Klingelbergstrasse 82, CH-4056 Basel, Switzerland}
\eads{\mailto{vdwivedi@thp.uni-koeln.de}}

\begin{abstract}
  We generalize the position- and momentum-space entanglement cuts to a family of cuts corresponding to regions in the classical phase space. We explicitly compute the entanglement spectra of free fermionic many-body wavefunctions for a family of phase space entanglement cuts that continuously interpolates between position- and momentum-space cuts. For inversion symmetric wavefunctions, the phase space entanglement spectrum possess a chiral symmetry, to which a topological index can be associated. 
\end{abstract}



\maketitle


\section{Introduction}
Quantum entanglement is arguably the most intriguing aspect of our current understanding of nature. Besides its implications for the ontology of quantum mechanics, it has also stimulated and enriched a diverse array of fields within theoretical physics\cite{calabrese2009entanglement} such as condensed matter\cite{amico2008entanglement, fidkowski_ent_ti, laflorencie_ent_rev}, quantum information\cite{horodecki_ent_rev,RevModPhys.82.1}, quantum field theory and string theory\cite{ryu2006holographic}. Insights from the study of quantum entanglement have led to the development of many analytical tools for characterization and simulation of extended many-body quantum systems, notable examples being topological entanglement entropy\cite{kitaev2006topological} and the density matrix renormalization group (DMRG)\cite{schollwock2005density}. 

In broad terms, quantum entanglement is the appearance of non-local correlations between local measurements on different parts of a system. Mathematically, this is encoded in the non-purity of the reduced density matrix, which can be quantified using the entanglement spectrum and associated entropies. In condensed matter systems, these measures of quantum entanglement have been used in diagnosing and classifying topological phases of matter\cite{fidkowski_ent_ti, ashvin_ent_inv, ryu2006entanglement, prodan2010entanglement, eisert2010colloquium}. Furthermore, for noninteracting fermionic systems, the entanglement spectrum can be interpreted as the spectrum of a free-fermion Hamiltonian, termed the \emph{Entanglement Hamiltonian}\cite{cheong2004many,peschel_corr, peschel_lattice}, which can be related to the edge Hamiltonian of the system\cite{fidkowski_ent_ti, ashvin_ent_inv}. 

A salient feature of entanglement is the dependence on the ``entanglement cut'', which often corresponds to a partition of a chosen basis of the single particle Hilbert space. In practice, one usually considers the eigenbasis of a Hermitian operator, or simultaneous eigenbases of a complete set of (mutually commuting) Hermitian operators, corresponding to some physically relevant quantities. A suitable choice of this basis can then lead to physical insights into the nature of the quantum correlations of the wavefunction under study. 

For quantum systems derived from the quantization of conventional classical mechanical systems, the most common choice is the configuration/position space basis\cite{prodan2010entanglement, pollmann_ent_spec, maria_ksl_ent, maria_ent_CI,  norm_ent_monte_carlo}. Computing the entanglement entropy or entanglement spectrum then entails tracing over modes localized in some chosen region in the position space (a \emph{position-space cut}). Alternatively, one may consider regions in the conjugate momentum space, i.e, a \emph{momentum-space cut}\cite{thomale2010nonlocal, mayukh-ian-taylor, balasubramanian_ent_rg,lundgren2014momentum,lundgren2016momentum}, which is highly nonlocal in the position-space picture. For systems with additional internal degrees of freedom, there are other possible entanglement cuts\cite{regnault2017entanglement, schoutens_laughlin_ent, schoutens_fqhe_ent} such as an orbital cut\cite{li2008entanglement} or particle cut\cite{sterdyniak2011extracting}, which reveal different aspects of entanglement in a wavefunction.

The dichotomy of position- and momentum-space entanglement cuts is representative of the dichotomy of position and momentum spaces in quantum mechanics, where one is typically the independent parameter in the wavefunction, while the other is a differential operator. In classical mechanics, on the other hand, the position and momentum spaces are treated on an equal footing as Lagrangian subspaces of the classical phase space.  
A degree of agnosticism in the choice of phase-space coordinates is further sanctioned by the modern geometric (coordinate-independent) formulation  of classical mechanics\cite{marsden2013introduction, arnold_book}, where the phase space is identified as a symplectic manifold, on which there exist an infinite multitude of valid choices for local position-momentum coordinates, following Darboux's theorem.
Unfortunately, quantization typically spoils this egalitarianism of phase-space coordinates. This is most explicit in geometric quantization\cite{woodhouse1997geometric}, where one must prescribe a \emph{polarization} on the prequantum line bundle, which, in physics terms, locally corresponds to choosing the wavefunction to be a function of position or momentum.

The perspective of viewing entanglement cuts in terms of the corresponding classical quantities raises a natural question: Can one define the usual entanglement measures for entanglement cuts \emph{along arbitrary directions} in the classical phase space, and if yes, what sorts of new features are encountered? In this paper we entertain these curiosities\footnote{ 
  Some of these ideas have previously been considered in Ref \cite{almeida2009entanglement}. 
}, 
using the Weyl-Wigner transform to switch between the classical phase space and the quantum Hilbert space. We define \emph{phase space entanglement cuts} for arbitrary codimension 1 hyperplanes in the phase space, which correspond to a highly nonlocal entanglement cut\cite{thomale2010nonlocal} in the real space. 

We study the entanglement spectra for a 1-dimensional free-fermionic many-body system, which can be computed in terms of quantities defined on the single-particle Hilbert space. The classical phase space is 2-dimensional, and we define an entanglement cut corresponding to lines in $\real^2$. In particular, we consider a family of entanglement cuts parametrized by $\theta \in [0,2\pi]$ interpolate continuously between the position- and momentum-space cuts. The resulting \emph{phase space entanglement spectrum}(PSES) can be interpreted as a 1-dimensional band structure. Furthermore, for inversion-symmetric wavefunctions, the PSES possesses a chiral symmetry, which can be used to associate a chiral invariant with certain many-fermion wavefunctions. In general, the PSES provides a complete classification of the many-fermion inversion symmetric wavefunctions. 

%



The rest of this article is organized as follows: In Section \ref{sec:ent}, we discuss the computation of entanglement spectra for free fermionic system. In Section \ref{sec:phsp}, we introduce the notion of entanglement cuts in phase space. In Section \ref{sec:rot}, we specialize to the family of phase-space cuts obtained by rotations in phase space for a 1D system, and derive convenient forms for the entanglement spectra as a function of rotation parameter. In Section \ref{sec:res}, we use this machinery to compute PSES for inversion symmetric wavefunctions, and define a winding number. We finally conclude in Section \ref{sec:conc}. Details of various computations are relegated to the appendices.

\section{Entanglement for noninteracting fermionic systems}    \label{sec:ent}
The natural habitat of quantum entanglement is a many-body Hilbert space $\hilbert$ that can be written as a tensor product of two subspaces, i.e, $\hilbert = \hilbert_\sA \otimes \hilbert_\sB$. Consider then a many body system described by a wavefunction $\ket{\Psi} \in \hilbert$, or equivalently by a pure density matrix $\hat\rho = \ket{\Psi}\bra{\Psi}$. Given the tensor product decomposition of $\hilbert$, usually termed an entanglement cut, the quantum entanglement is encoded in the reduced density matrix $\hat\rho_\sA = \Tr_\sB \hat\rho$, where the trace is taken over $\hilbert_\sB$. This can be arrived at by a Schmidt decomposition of the wavefunction as
\beq 
  \ket{\Psi} = \sum_\alpha \sqrt{\lambda_\alpha} \ket{A_\alpha} \otimes \ket{B_\alpha} \implies 
  \hat\rho_\sA = \sum_\alpha \lambda_\alpha \ket{A_\alpha} \bra{A_\alpha}, 
\eeq 
where $\lambda_\alpha \in [0,1]$ are the Schmidt eigenvalues, and $\{\ket{\sA_\alpha} \}_\alpha, \{\ket{\sB_\alpha} \}_\alpha$ are orthonormal in $\hilbert_{\sA,\sB}$, respectively. Thinking of $\hat\rho_\sA$ as a thermal Gibbs density matrix for a system described by the Hamiltonian $\hlt_\sE$, i.e, $\hat\rho_\sA \propto \rme^{-\hlt_\sE}$, the \emph{entanglement spectrum} is defined as the spectrum of $\hlt_\sE$, i.e, $\ve_{\sE, \alpha} = \ve_0 - \ln \lambda_\alpha$ (up to an irrelevant constant $\ve_0$). Futher measures of entanglement, such as entanglement entropies, can then be computed from the entanglement spectrum. 

In this article, we are interested in entanglement in many-body system consisting of free fermions, which naturally occur as eigenstates of noninteracting fermionic Hamiltonians. The entanglement spectrum for such states can be computed using the single particle wavefunctions either from the fermionic correlation functions\cite{cheong2004many,peschel_corr, peschel_lattice} or the overlap matrix in the reduced subsystem\cite{peschel_spheroidal, klich_entropy_bound}. Since the latter approach is more amenable to generalization to phase space cuts, we describe it in some detail in the following.

The many-body states of free fermions are described by Slater determinants over single particle wavefunctions. More formally, the total Hilbert space decomposes as  $\hilbert = \bigoplus_{n = 0}^\infty \hilbert_n$, where $\hilbert_n$ is the $n$-particle Hilbert space, defined as an antisymmetrized tensor product over $n$ copies of the single particle Hilbert space $\hilbert_1$ (See \ref{app:hilbert} for more details). Explicitly, given a set of $N$ single-particle wavefunctions $\ket{\psi_a} \in \hilbert_1, \, a = 1, \dots N$, we can form a $N$-body state as
\beq 
  \ket{\Psi} = \ket{\psi_1} \wedge \ket{\psi_2} \wedge \dots \wedge \ket{\psi_N} 
  = \psi_1^\dg \psi_2^\dg \dots \psi_N^\dg \ket{\Omega},
\eeq 
where $\ket{\Omega} \in \hilbert_0 \cong \cmplx$ is the number vacuum state. Since the density matrix $\hat\rho = \ket{\Psi}\bra{\Psi}$ is invariant under $\U(N)$ rotations among the single particle wavefunctions, we set $\braket{\psi_a}{\psi_b} = \delta_{ab}$ in the following without loss of generality. 

Choosing a subspace $\hilbert_{1,\sA} \subset \hilbert_1$ such that $\hilbert_1 = \hilbert_{1,\sA} \oplus \hilbert_{1,\sB}$, the total Hilbert space can naturally be written as an antisymmetrized tensor product $\hilbert = \hilbert_\sA \wedge \hilbert_\sB$, where $\hilbert_{\sA/\sB}$ are constructed from antisymmetric tensor products of $\hilbert_{1,\sA/\sB}$ (See \ref{app:hilbert} for more details). We also define the vacuua $\hilbert_{0, \sA/\sB} \cong \cmplx$ so that $\hilbert_0 = \hilbert_{0,\sA} \wedge \hilbert_{0, \sB}$. 
Defining the orthogonal projectors $\proj_{\sA/\sB} \cn \hilbert_1 \to \hilbert_{1,\sA/\sB}$, the single particle wavefunctions can be written as
\beq 
  \ket{\psi_a} = \left\{ \proj_{\sA} \ket{\psi_a}  \right\} \otimes \ket{\Omega_\sB} + \ket{\Omega_\sA} \otimes \left\{ \proj_{\sB} \ket{\psi_a} \right\},    \label{eq:wf_decom}
\eeq 
where $\ket{\Omega_{\sA/\sB}} \in \hilbert_{0,\sA/\sB}$ are the vacuum states. This splits the many-body density matrix into a sum over terms with separated $\hilbert_\sA$ and $\hilbert_\sB$ contributions. 

To perform the trace over $\hilbert_\sB$, we need to construct an orthonormal basis for $\text{span}\{\proj_{\sB} \ket{\psi_a} \} \subset \hilbert_{1,\sB}$. To that end, consider the overlap (Gramian) matrix defined as:
\beq 
   \overlap_{ab} = \braket{\psi_a}{\psi_b}_\sA \equiv \braket{\proj_\sA \psi_a}{\proj_\sA \psi_b},     \label{eq:overlap_def}
\eeq 
where $\langle , \rangle_\sA$ denote the inner product on $\hilbert_{1,\sA}$. Since $\braket{\psi_a}{\psi_b} = \delta_{ab}$, the overlap matrix on $\hilbert_{1,\sB}$ is simply $\id - \overlap$. Diagonalizing $\overlap$, we get
\beq 
   \overlap = \uop^\dg \, \Theta \, \uop, \qquad \Theta = \diag{ \eig_1, \dots \eig_N }, \; \eig_a \in [0,1],
\eeq  
where the bounds on $\eig_a$ follow from the properties of Gramian matrices(\ref{app:gram}). Assuming for the moment that $\mu_a \neq 0,1$, we define the normalized wavefunctions 
\beq 
  \ket{A_a} = \frac{1}{\sqrt{\eig_a}} \sum_{b=1}^N \uop_{ab}^\ast \, \proj_{\sA} \ket{\psi_b}, \qquad 
  \ket{B_a} = \frac{1}{\sqrt{1-\eig_a}} \sum_{b=1}^N \uop_{ab}^\ast \, \proj_{\sB} \ket{\psi_b}, 
\eeq 
on $\hilbert_{\sA, \sB}$, which are orthonormal, since 
\beq 
  \fl \qquad \braket{A_a}{A_{a'}}_\sA 
  = \frac{1}{\sqrt{\eig_a \eig_{a'}}} \sum_{bb'} \uop_{ab}^\pdg \, \uop^\ast_{a'b'} \braket{\psi_b}{\psi_{b'}}_\sA
  = \frac{1}{\sqrt{\eig_a \eig_{a'}}} \left[\uop \, \overlap \, \uop^\dg \right]_{aa'} 
  = \delta_{aa'}, 
\eeq 
and a similar computation for $\ket{B_a}$. 
After a unitary rotation by the $\uop$'s in the space of single particle wavefunctions, using the invariance of the Slater determinant under $\U(N)$ rotations, the many body wavefunction becomes
\beq 
  \ket{\Psi} 
  = \bigwedge_{a=1}^N \left[ \sqrt{\eig_a} \, \ket{A_a} \otimes \ket{\Omega_\sB} + \sqrt{1-\eig_a} \, \ket{\Omega_\sA} \otimes \ket{B_a} \right].   \label{eq:modes_orth}
\eeq 
This expression also works if $\eig_a = 0,1$, since for those cases, we do not need to define the corresponding $\ket{A_a}$ or $\ket{B_a}$, respectively. 

The many-body density matrix can then be expanded as 
\beq 
  \hat\rho =  \sum_{N_\sA=0}^N \left[ \sum_{\va, \vb} \left( \prod_{i=1}^{N_\sA} \eig_{a_i} \right) \left( \prod_{i=1}^{N-N_\sA} (1-\eig_{b_i}) \right) \ket{\Psi^\sA_\va} \bra{\Psi^\sA_\va} \otimes \ket{\Psi^\sB_\vb} \bra{\Psi^\sB_\vb}  \right],
\eeq 
where 
\beq 
  \ket{\Psi^\sA_\va} =  \bigwedge_{i=1}^{N_\sA} \ket{A_{a_i}}, \qquad 
  \ket{\Psi^\sB_\vb} =  \bigwedge_{i=1}^{N-N_\sA} \ket{B_{b_i}},  
\eeq 
where $\va \subset \{1, \dots N \}$ with $N_\sA$ elements, and $\vb = \{1, \dots N\} \backslash \va$. The trace over $\hilbert_\sB$ now gets rid of $\ket{\Psi^\sB_\vb}$'s, resulting in the reduced density matrix:
\beq 
  \hat\rho_{\sA, k} = \det(\id-\overlap) \sum_{\va} \left( \prod_{i=1}^{N_\sA} \frac{\eig_{a_i}}{1-\eig_{a_i}} \right) \ket{\Psi_\va} \bra{\Psi_\va}, 
\eeq
where we have used the fact that $\det(\id-\overlap) = \prod_{i=1}^N \left( 1 - \eig_{i} \right)$. The reduced density matrix can be concisely written as\cite{klich_entropy_bound} the $2^N \times 2^N$ diagonal matrix
\beq 
 \hat\rho_\sA = \bigotimes_{a=1}^N \left( 
  \begin{array}{cc}
    \eig_a & 0 \\ 
    0 & 1-\eig_a
  \end{array}  
  \right),   \label{eq:redc_dens_mat}
\eeq 
Physically, Equation \ref{eq:modes_orth} represents a Slater decomposition over single particle modes, which are linear superpositions of orthonormal modes supported only in $\hilbert_{1,\sA}$ and those supported only in $\hilbert_{1,\sB}$, with the corresponding probabilities $\eig_a$ and $1-\eig_a$, respectively. 

For noninteracting fermionic systems, the single particle sector of the reduced density matrix is of particular interest, since its knowledge can be used to reconstruct the many-body reduced density matrix\cite{peschel_corr}. For these eigenvalues, we define the entanglement energies as 
\beq 
  \lambda_a = \frac{\eig_a}{1-\eig_a} \det(\id-\overlap) \implies 
  \ve_{E, a} = -\ln \left( \frac{\eig_a}{1-\eig_a} \right),   \label{eq:ent_eng_def} 
\eeq 
where we have ignored $\ve_0 = \Tr \ln (\id-\overlap)$, since $\ve_{\sE, a}$ is defined only up to a constant. The entanglement energies in the $k$-particle sector, again ignoring $\ve_0$, is then simply given by all possible sums of $k$ single-particle entanglement energies. Thus, the entanglement spectrum can be interpreted as the energies of noninteracting fermionic many-body states constructed from the eigenstates of the single-particle Hamiltonian
\beq 
  \hlt_\sE = -\ln \left[ \overlap \left( \id-\overlap \right)^{-1} \right] = \ln \left[ \overlap^{-1} - \id \right],   \label{eq:ent_hlt}
\eeq
which is the entanglement Hamiltonian. Clearly, for free fermions, the knowledge of the overlap matrix is sufficient to compute the entanglement spectrum. Thus, in the next sections, we shall simply compute this overlap matrix as a function of the phase space entanglement cut.

%


\section{Entanglement Cuts in Phase Space}    \label{sec:phsp}
In this section, we clarify the meaning of a \emph{phase space entanglement cut} in terms of the $d$-dimensional single particle Hilbert space\footnote{
  Note that the discussion of Section \ref{sec:ent} does not allude to the contents of the single particle Hilbert space. In particular, the computations readily generalize to fermions with internal (spin/orbit) degrees of freedom, simply by taking the corresponding inner product in Equation \ref{eq:overlap_def}. However, we only consider cases without any internal degrees of freedom in this article. 
}
$\hilbert_1 = L^2(\real^d)$. A direct connection between operations in the classical phase space $\real^{2d}$ and the unitary transforms on the single particle states is provided by the Wigner-Weyl transformation, which is used to define entanglement cuts corresponding to arbitrary hyperplanes in the phase space.

\subsection{Choosing an entanglement cut}        \label{sec:phsp_cut}
As discussed in Section \ref{sec:ent}, a choice of bipartition of the single particle Hilbert space decomposes the many-body Hilbert space into a tensor product structure, for which one can compute the entanglement spectrum. In practice, this bipartition of $\hilbert_1$ is most conveniently defined by choosing a basis of $\hilbert_1$ and defining the ``subsystem'' $\sA$ as the span of a subset of the basis vectors. For instance, for fermions on a $d$-dimensional space $\real^d$, a convenient basis of $\hilbert_1$ is the position basis, \viz, $\{ \ket{\vx}, \; \vx \in \real^d \}$. One can then choose a region $\sA \subset \real^d$ and define $\hilbert_{1,\sA} \equiv \text{span}\left\{ \ket{\vx}, \, \vx \in \sA \right\}$. Another convenient basis is the momentum basis $\{ \ket{\vp}, \; \vp \in \real^d \}$, using which one can define a momentum-space cut. 

Since these bases correspond to the same vector space, they are be related by a unitary transformation. Indeed, this is just the Fourier transform
\beq 
  \ket{\vp} \equiv \int_{\real^{d}} \frac{\rmd^dx}{(2\pi)^{d/2}} \, \rme^{i \vp\cdot\vx} \ket{\vx}, \quad \vp \in \real^{d}.	
\eeq
These two bases are in direct correspondence with the position and momentum in classical mechanics. However, classically, one thinks of the phase space $\real^{2d}$ with coordinates $\bxi = (\vx, \vp)$, and the $\vx$-axis can be mapped into the $\vp$-axis by a rotation $ g \in \SO(2d)$. Thus, for $\vx\to\vp$, this coordinate transformation in the classical phase corresponds to a unitary transformation in the quantum Hilbert space. Since the coordinates $\vx$ and $\vp$ are treated on an equal footing and one has a plethora of transformations associated with the phase space, it is natural to attempt to associate unitary operators on $\hilbert_1$ with those transformations.

To make the connection with entanglement cuts clearer, we can also associate regions of the phase space with choices of subsystems. For instance, a position space cut defined by a choice of subsystem $\sA \subset \real^d$ can be ``associated with'' the region $\sA' \equiv \sA \times \real^d$ in the phase space, since this entanglement cut  puts no constraints on the momentum $\vp$. More explicitly, consider the position space cut, defined by the choice of a subsystem
\beq
  \sA_\vx = \{\vx\in \real^d : x_1 \geqslant 0 \},     \label{eq:pos_cut}
\eeq 
with the $(d-1)$ dimensional cut plane $\Pi \subset \real^d$ corresponding to $x_1 = 0$. The corresponding phase space partition $\wt{\sA}_\vx$ and the $(2d-1)$ dimensional cut plane $\wt{\Pi}_\vx$ can be explicitly written as
\beq 
  \wt{\sA}_\vx = \{\bxi \in \real^{2d} : \xi_1 \geqslant 0 \}, \qquad 
  \wt{\Pi}_\vx = \{\bxi \in \real^{2d} : \xi_1=0\}.      \label{eq:pos_cut_phsp}
\eeq
Mathematically, the overlaps for the entanglement cut $\sA_\vx$ turn out to be equal to integrals of a corresponding \emph{Wigner function} over $\wt{\sA}_\vx$, as discussed in more detail in Section \ref{sec:phsp_wig}. 

One particular allure of this approach lies in the fact that the classical phase space offers many continuous families of coordinate transformations, whose quantum equivalents are not immediately obvious. For instance, the coordinate axes can be continuously rotated into the momentum axes by a set of rotations in $\SO(2d)$. We next discuss the set of transformations that leave the phase space invariant, and construct the corresponding unitary operators on the single particle Hilbert space. 

\subsection{Classical mechanics and families of entanglement cuts}       \label{sec:phsp_cm}
We begin by recalling a few facts about the modern approach to classical mechanics\cite{marsden2013introduction}, which relies on the symplectic structure of the phase space. More explicitly, the classical phase space is equipped with a closed, nondegenerate (symplectic) 2-form $\omega$, which can be written in canonical coordinates $(\vx, \vp)$ on $\real^{2d}$ as
\beq 
  \omega \equiv \frac{1}{2} \omega_{ij} d\xi^i \wedge d\xi^j = \rmd\vp \wedge \rmd\vx =  \rmd p_i \wedge \rmd x^i. 
\eeq 
We also define $\bom$ is the $2d$-dimensional antisymmetric matrix with entries $\omega_{ij}$, so that in canonical coordinates,
\beq 
  \bxi = \vecenv{\vx}{\vp}, \qquad 
  \bom = 
  \left(\begin{array}{cc} 
	   0 & \id_d  \\ 
	    -\id_d & 0
  \end{array}\right).
\eeq
The symplectic form also defines the Poisson bracket, which can be quantized by the Dirac's prescription to obtain the canonical commutation relations $[\hat{x}_j ,\hat{p}_k] = i \delta_{jk} \hat{1}$. Explicitly, 
\beq 
  \{\xi^j, \xi^k\} = \left(\bom^{-1} \right)^{jk} \longrightarrow [\hat\xi^j ,\hat{\xi}^k] = i \left(\bom^{-1} \right)^{jk} \hat{1}, 
\eeq 
where $\hat{\bxi}$ is a Hermitian operator on $L^2(\real^d)$.

The set of transformations of the phase space that leave the symplectic form invariant consist of translations, rotations and dilatations in the phase space $\real^{2d}$. These form the Lie group of linear canonical transformations, also known as the inhomogeneous symplectic group\cite{burdet1978generating,low2012projective} $\ISp(2d,\real) \equiv \real^{2d} \rtimes \text{Sp}(2d,\real)$. Here, $\real^{2d}$ is the group of translations, i.e, the abelian Lie group formed by $\real^{2d}$ with vector addition as the group composition law. Explicitly, the Lie group consists of the $(2d+1)\times(2d+1)$ matrices of the form 
\beq 
  g = 
    \left(\begin{array}{cc} 
	    \vA & \vb  \\ 
	    0 & 0
  \end{array}\right), \qquad 
  \vA \in \Sp(2d, \real), \quad \vb \in \real^{2d}.
\eeq
which act on the column vector $(\bxi^T, 1)^T$ as $\bxi \mapsto \fA \, \bxi + \fb$. The corresponding Lie algebra $\mathfrak{isp}(2d,\real)$ consists of matrices of the form 
\beq
  X =
  \left(\begin{array}{cc} 
	    \fA & \fb  \\ 
	    0 & 1
  \end{array}\right), \qquad 
  \fA \in \mathfrak{sp}(2d, \real), \quad \fb \in \real^{2d},    \label{eq:isp_gen}
\eeq 
where $\mathfrak{sp}(2d, \real)$ and $\real^{2d}$ are the Lie algebras of $\Sp(2d, \real)$ and $\real^{2d}$, respectively. The generators of this Lie algebra correspond precisely to the quadratic Hamiltonians\cite{burdet1978generating}. In \ref{app:isp}, we show that these Hamiltonians are explicitly given by 
\beq 
  H(\bxi) = \frac{1}{2} \bxi^T  \bom \fA \, \bxi + \fb^T \bom \bxi,
\eeq 
for the generator defined in Equation \ref{eq:isp_gen}.

In the next subsections, we associate a unitary operator $\uop_g \in U(\hilbert_1)$ with each $g \in \ISp(2d,\real)$ by explicitly defining a (projective) unitary representation of $\ISp(2d,\real)$ on $\hilbert_1$. For the moment, assuming the existence of these unitaries, we explicitly discuss certain families of entanglement cuts corresponding to one parameter subgroups $g(t) \equiv \exp(tX) \in \ISp(2d,\real)$ for $t \in \real$ and $X \in \mathfrak{isp}(2d,\real)$. These examples illustrate the connection between the phase space picture and the conventional entanglement cut pictures. Explicitly, we consider the entanglement cuts corresponding to the region
\beq  
  \wt{\sA}_g = \{ \bxi \in \real^{2d} : \left[ g^{-1} \bxi \right]_1 \geqslant 0 \}, \qquad 
  \wt{\Pi}_g =  \{ \bxi \in \real^{2d} : \left[ g^{-1} \bxi \right]_1 = 0 \}      \label{eq:phsp_reg_def}
\eeq 	
for certain $g(t) \in \ISp(2d, \real)$, and we have taken a cut along the first coordinate for convenience. Here, we interpret the transformation $\ket{\psi} \to \uop_g \ket{\psi}$ as an \emph{active} transformation, while $\bxi \to g^{-1} \bxi$ is the corresponding \emph{passive} transformation. We also discuss the Hamiltonians generating these one-parameter subgroups, whose (Weyl-)~quantization would turn out to be generators of the corresponding transformation in phase space, as we shall explicitly see for a special case in Section \ref{sec:rot}.


\begin{enumerate}
\item {\bf Phase-Space Translations:}
These transformations replicate the conventionally studied families of position- and momentum-space cuts. Consider a general phase space translation $ \bxi \mapsto \bxi + \mathbf{u}t$, where $t \in \real$ and $\mathbf{u}\equiv(\mathbf{u}_x,\mathbf{u}_p)$ is a unit vector in $\real^{2d}$. The one parameter subgroup of $\ISp(2d, \real)$ and the corresponding Lie algebra generators are given by
\beq
  g(t) = 
    \left(\begin{array}{cc} 
	    \id & \mathbf{u}\, t \\ 
	    0 & 1
    \end{array}\right) 
    \implies     
    X =  
    \left(\begin{array}{cc} 
	    0 & \mathbf{u} \\ 
	    0 & 0
    \end{array}\right),
\eeq
and the corresponding Hamiltonian is 
\beq
  H(\bxi) = \vu^T \bom \, \bxi = \mathbf{u}_x \cdot \vp - \mathbf{u}_p \cdot \vx.  
\eeq
To further unpack the meaning of these cuts, consider $\vu_x = \{1, 0, \dots\}$ and $\vu_p = \nullv$. The entanglement cuts can explicitly be written as 
\beq 
  \wt{\sA}_t = \{ \bxi \in \real^{2d} : x_1 - t \geqslant 0 \} \implies \sA_t = \{ \vx \in \real^d : x_1 \geq t \},
\eeq
which denotes the familiar family of position space cuts along $x_1$. A similar argument for $\vu$ along a momentum direction (alongwith a rotation) leads to a family of momentum space cuts.

\item {\bf Phase Space Rotations:}
These result in a family of entanglement cuts that continuously interpolates between the position- and momentum-space cuts.  Explicitly, consider the one parameter subgroup that implements $\text{SO}(2)$ rotations among each of the coordinate pairs $(x_k,p_k)$ for $k = 1,\ldots ,d$, i.e, 
\beq 
  \fl \qquad
  \vecenv{x_k}{p_k} \mapsto 
    \left(\begin{array}{cc} 
	    \cos t & -\sin t \\ 
	    \sin t & \cos t
    \end{array}\right)  \vecenv{x_k}{p_k}
    = 
    \exp \left[ -t \left(\begin{array}{cc} 
	    0 & 1 \\ 
	    -1 & 0
    \end{array}\right) \right] \vecenv{x_k}{p_k},
\eeq 
where the matrix is the symplectic matrix $\bom$ restricted to the $x_k$--$p_k$ subspace. Thus, the general transformation on $\real^{2d}$ is $\bxi \mapsto e^{i \bom t} \bxi$, so that 
\beq
  g(t) = 
  \left(\begin{array}{cc} 
	    \exp(-\bom t) & 0 \\ 
	    0 & 1
  \end{array}\right) 
    \implies     
    X =  
    \left(\begin{array}{cc} 
	    -\bom & 0 \\ 
	    0 & 0
    \end{array}\right).
\eeq
Remarkably, the corresponding Hamiltonian generator is 
\beq
  H(\bxi) = -\frac{1}{2} \bxi^T  \bom^2 \bxi = \frac{1}{2} \bxi^T \bxi =  \frac{1}{2} \left( \vx^2 + \vp^2 \right),    \label{eq:rot_hlt_def}
\eeq
which is just the simple harmonic oscillator Hamiltonian in $d$-dimensions. For $t=0$, we recover the position space cut along $x_1 = 0$, while for $t = \pi/2$, we get a momentum space cut along $p_1 =0$. Finally, for $t=\pi$, we get 
\beq 
  \wt{\sA}_{\pi} = \{ \bxi \in \real^{2d} : [-\bxi]_1 \geqslant 0 \} \implies \sA_{\pi} = \{ \vx \in \real^d : x_1 \leq 0 \},
\eeq
which corresponds to an inversion about the origin! Thus, for the case at hand, a phase space rotation by $\pi/2$ reproduces the original system with the two subsystems swapped, so that one recovers the original entanglement spectrum. This has interesting consequences for the entanglement spectra of inversion symmetric many-body wavefunctions, as discussed in Section \ref{sec:res}. 



\item  {\bf Equal Area Shear:} 
Finally we consider an equal area shear in each of the coordinate pairs $(x_k,p_k)$ for $k = 1,\ldots ,d$, i.e, 
\beq 
  \fl \qquad
  \vecenv{x_k}{p_k} \mapsto 
    \left(\begin{array}{cc} 
	    \rme^t & 0 \\ 
	    0 & \rme^{-t}
    \end{array}\right)  \vecenv{x_k}{p_k}
    = 
    \exp \left[ t \left(\begin{array}{cc} 
	    1 & 0 \\ 
	    0 & -1
    \end{array}\right) \right] \vecenv{x_k}{p_k},
\eeq 
Thus, the general transformation on $\real^{2d}$ is $\bxi \mapsto e^{i \vS t} \bxi$ with the block matrix $\vS = \text{diag} \left\{ \id_{d}, -\id_{d} \right\}$, so that 
\beq
  g(t) = 
  \left(\begin{array}{cc} 
	    \exp(\vS t) & 0 \\ 
	    0 & 1
  \end{array}\right) 
    \implies     
    X =  
    \left(\begin{array}{cc} 
	    \vS & 0 \\ 
	    0 & 0
    \end{array}\right), 
\eeq
and the corresponding Hamiltonian generator is 
\beq
  H(\bxi) = \frac{1}{2} \bxi^T  \vS \bom \bxi 
  = -\frac{1}{2} \bxi^T  \left(\begin{array}{cc} 
	    0 & \id \\ 
	    \id & 0
    \end{array}\right)   
  \bxi =  -\vx \cdot \vp.
\eeq
The meaning of this transformation is revealed if instead of the cut defined in Equation \ref{eq:phsp_reg_def}, we start off with a momentum space cut defined as 
\beq
  \sA = \{ \vp \in \real^d : |\vp| \leqslant  \Lambda \}, \qquad \wt{A} = \real^d \times \sA.
\eeq
Then, applying $\uop_{g(t)}$ on a density matrix $\hat\rho$ is equivalent to rescaling the ``cutoff'' $\Lambda \mapsto \Lambda \e^{t}$, and tracing over $\hilbert_\sB$ represents a coarse-graining of high momentum states. Together, these constitute a single iteration in a \emph{renormalization group} transformation on the density matrices\cite{balasubramanian_ent_rg}. 


\end{enumerate}


\subsection{Wigner-Weyl transforms and Wigner functions}      \label{sec:phsp_wig}
In order to explicitly define a projective unitary representation of $\ISp(2d, \real)$ on the single particle Hilbert space, we turn next to a phase-space formulation\cite{peres2006quantum,folland2016harmonic,deGosson2017harmonic} of quantum mechanics. This approach maps density matrices, and quantum mechanical operators in general, into functions in the classical phase space. Many such (formally equivalent) representations of quantum mechanics have been proposed\cite{cohen1966phsp}, which are useful in different setups, a few examples being the Wigner-Weyl transformation(WWT), the Glauber–Sudarshan P-representation and the Husimi Q-representation. 

In this article, we consider the Wigner-Weyl transform (WWT)\cite{moyal, case_wigner_rev}, a linear bijection\footnote{
  In this work we will only deal with trace-class operators and smooth Weyl symbols where the WWT is a well behaved isomorphism. The issue of the regularity of $O(\bxi)$ and $\hat{O}$ is discussed more comprehensively in the literature\cite{folland2016harmonic, voros1978algebra}. 
}
between quantum mechanical operators and phase-space functions, so that given a linear operator $\hat{O} \cn \hilbert_1\to \hilbert_1$, the WWT yields a function $O \cn \real^{2d} \to \real$, termed the \emph{Weyl symbol} of $\hat{O}$. Explicitly, the WWT and its inverse--the Weyl quantization prescription--are defined as 
\beq
  O (\bxi) = \Tr \{\hat{O} \; \wigE(\bxi)\} \Longleftrightarrow 
  \hat{O} = \int_{\real^{2d}} \rmd\mu(\bxi) \, O(\bxi) \wigE(\bxi),   \label{eq:wigner_def}
\eeq 
where the kernel of the transform and the volume measure on the phase space are 
\beq 
  \wigE(\bxi) = \int d\mu(\bxi') \,  \rme^{i \bom(\bxi, \bxi')} \; \wige(\bxi'), \qquad 
  \rmd\mu(\bxi) \equiv \left( \frac{\omega}{2\pi} \right)^d = \frac{\rmd^d x \, \rmd^dp}{(2\pi)^{d}},
\eeq
respectively, and 
\beq
  \wige(\bxi') \equiv  \rme^{i \bom(\bxi', \hat{\bxi})}
  = \rme^{i \left(\vx' \cdot\hat{\vp} - \vp' \cdot \hat{\vx} \right)}
  = \rme^{- \frac{i}{2} \vx' \cdot \vp'} \rme^{ - i \vp'\cdot \hat{\vx}} \rme^{i \vx' \cdot \hat{\vp}},    
  \label{eq:def_wige}
\eeq
where we have defined $\hat{\bxi} \equiv (\hat\mathbf{{x}},\hat{\vp})$ and  $\bom(\bxi,\bxi') \equiv \bxi^T \bom \bxi'$, and used the Zassenhaus formula in the last step. 

Of particular importance is the Weyl symbol of the a density matrix $\hat{\rho}$, commonly termed the \emph{Wigner function}\cite{moyal} $W(\bxi)$. Since $\Tr\hat\rho = 1$ by normalization, the Wigner function also integrates to one over the phase space\footnote{
  Since the WWT is an isometry on trace-class operators.
}.
However, it cannot be interpreted as a probability distribution, since $W(\bxi)$ can take negative values. This is not a ``bug''; rather, it is simply a manifestation of the superposition principle of quantum mechanics. More precisely, since the WWT is linear, a linear superposition of wavefunctions corresponds to the an addition of the corresponding Wigner functions, so that the negative values of the Wigner function takes care of the possible destructive interference between wavefunctions.

For pure density matrices $\hat{\rho} =|\psi\rangle \langle \psi|$, the Wigner function $W(\bxi) = \bra{\psi} \wigE(\bxi) \ket{\psi}$ can be expressed in terms of the position space wavefunction $\psi(\vx) = \braket{\vx}{\psi}$  as 
\beq
  W(\vx,\vp) = \int \rmd^d x' \,  \rme^{- i \vp\cdot \vx'} \psi^\ast\left( \vx - \frac{\vx'}{2} \right) \psi\left( \vx + \frac{\vx'}{2} \right),     
\label{eq:wig_def_wf}
\eeq
which is sometimes more convenient for explicit calculations. Its inverse is in turn 
\beq
  \hat\rho = \int \rmd^d x_1 \rmd^d x_2  \left[ \int \frac{\rmd^d p}{(2\pi)^d}  \,  W\left(\frac{\vx_1 + \vx_2}{2}, \vp\right) \,  \rme^{i \vp\cdot (\vx_1 - \vx_2)} \right] \ket{\vx_1}  \bra{\vx_2}  
\label{eq:wigner_inv},
\eeq
from which the single particle wavefunction can be read off by the definition of $\hat\rho$. These expressions are explicitly derived from the defintion of the WWT in \ref{app:wigner}. 

From Equation \ref{eq:wig_def_wf}, we also note that the marginal distributions obtained by integration over either position or momentum coordinates produces the correct probability distributions in the remaining coordinate,  i.e, 
\beq 
  \fl \qquad 
  \int_{\vp \in \real^d} \frac{\rmd^dp}{(2\pi)^d} W(\vx, \vp) =  |\psi(\vx)|^2, \qquad
  \int_{\vx \in \real^d} \frac{\rmd^dx}{(2\pi)^d} \, W(\vx, \vp) =  |\wt{\psi}(\vp)|^2,
\eeq 
where $\wt{\psi}(\vp) = \braket{\vp}{\psi}$ is the Fourier transform of $\psi$. For a mixed density matrix, we also have off-diagonal terms of the form $\ket{\psi}\bra{\phi}$, for which we define the cross-Wigner function
\beq 
  \wt{W}(\vx,\vp) \equiv \bra{\phi} \wigE(\vx,\vp) \ket{\psi} = \int_\real \rmd^d x' \,  \rme^{- i \vp \cdot \vx'} \phi^\ast \left( \vx - \frac{\vx'}{2} \right) \psi\left( \vx + \frac{\vx'}{2} \right),
\eeq 
which can in general be complex. Its marginal distribution can be used to compute the overlap 
\beq 
  \int_{\vp \in \real^d} \frac{\rmd^dp}{(2\pi)^d} \, \wt{W}(\vx, \vp) = \varphi^\ast(\vx) \psi(\vx),    \label{eq:wig_overlap}
\eeq 
which can be used to compute half space overlaps, by integrating the Wigner function over $\sA \times \real$. Thus, the phase space cuts introduced in Section \ref{sec:phsp_cut} can be visualized as the regions in the phase space over which one needs to integrate the relevant Wigner function to obtain the overlap matrix, which can be used to compute the entanglement spectrum.

\subsection{The projective unitary representation of $\ISp(2d, \real)$}    \label{sec:phsp_rep}
\begin{figure}[ht]
  \centering
  \includegraphics[width=0.85\textwidth]{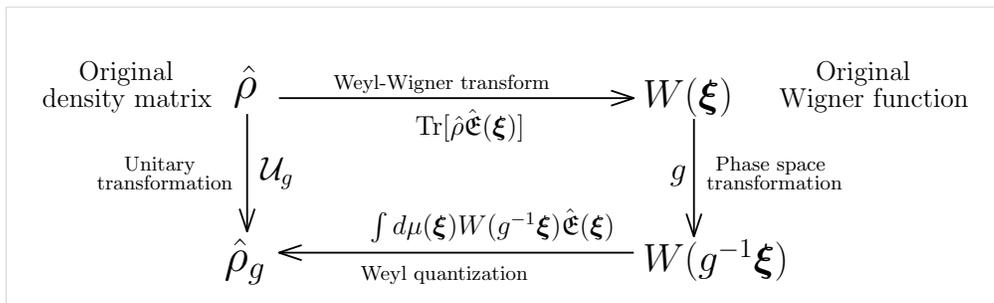} 
  \caption{Schematic for the implementation of a $\ISp(2d, \real)$ on a single-particle density matrix via the Wigner-Weyl transformation.} 
  \label{fig:flowchart}
\end{figure}

The WWT provides us with an indirect route of implementing $\ISp(2d, \real)$ transformations on single particle density matrices. Since the Wigner function transforms as a scalar under the transformations of the phase space, given a density matrix, we can compute its Wigner function, implement the requisite phase space transformation, and then use Weyl quantization to obtain the resulting density matrix. This results in a unitary transformation (see Equation \ref{eq:unit_xform}) on the space of density matrices, which defines the wavefunctions only up to a phase, so that one gets a projective unitary representation of $\ISp(2d, \real)$ on $\hilbert_1$. The sequence of operations is schematically depicted in Fig~\ref{fig:flowchart}. 

More formally, $\forall \, g \in \ISp(2d,\real)$, there is an induced action
\beq
  g^*:W(\bxi) \mapsto W_g(\bxi) = W \left(g^{-1} \bxi \right),
\eeq
which lifts to an induced action on the density matrix 
\beq\label{eq:rho_g}
  g^*: \hat\rho \mapsto \hat\rho_g = \int \rmd\mu(\bxi) \; W(g^{-1} \bxi)\; \wigE(\bxi),   \label{eq:ind_rep}
\eeq
which is a unitary transformation\cite{folland2016harmonic, deGosson2017harmonic} on $\hilbert_1 \otimes \hilbert_1^\ast$. One way to see the unitarity is to consider $\mathbb{V}$, the vector space of linear trace class operators on $\hilbert_1^\ast \otimes \hilbert_1$, which form a Hilbert space under the Hilbert-Schmidt inner product, defined as $\langle \hat{A}, \hat{B} \rangle \equiv \Tr ( \hat{A}^\dg \hat{B} )$. The set of operators $\{ \wigE(\bxi), \, \bxi \in \real^{2d} \}$ form a basis of this space, so that WWT can be thought of simply as expansion of an operator $\hat{A}$ in this basis, with $O(\bxi) \in \real$ being the coefficients. The completeness of this basis is equivalent to the statement that the WWT is a bijection. Thus, we can rewrite Equation \ref{eq:ind_rep} more explicitly as 
\beq 
  \hat\rho_g = \int \rmd\mu(\bxi) \; \Tr \{ \hat\rho \; \wigE(g^{-1} \bxi)\} \wigE(\bxi) = \int \rmd\mu(\bxi) \; \Tr \{ \hat\rho \; \wigE(\bxi)\} \wigE(g \, \bxi),    \label{eq:unit_xform}
\eeq
where we have used the fact that the symplectic form, and hence the phase space measure, is invariant under $\ISp(2d, \real)$, so that $\rmd\mu(g \, \bxi) = \rmd\mu(\bxi)$. This equation can then be interpreted as a basis transformation on $\mathbb{V}$. But since all orthonormal bases are related to one another by a unitary transform, we can deduce that $\hat\rho \mapsto \hat\rho_g$ is a unitary transform over $\mathbb{V}$. 

In principle, given a wavefunction $\ket{\psi} \in \hilbert_1 \cong L^2(\real^n)$ and a transformation $g \in \ISp(2d, \real)$, one can explicitly go through this procedure to obtain the transformed wavefunction $\ket{\psi_g}$. However, for a given family of transformations $g(t)$, it is more convenient to construct an explicit operator (typically as an integration kernel) that implements this operation on the real-space wavefunction (See, for instance, Section 5 of Ref \cite{deGosson2017harmonic}). In the next section, we explicitly construct such an integration kernel for the phase space rotations.

\section{Phase space rotation}   \label{sec:rot}
We now restrict to 1-dimensional systems, so that the phase space is $\real^2$. We show that the unitary representation of the rotation subgroup $\SO(2) \subset \ISp(2,\real)$ on the single-particle Hilbert space $\hilbert_1 \cong L^2(\real)$ is a \emph{fractional Fourier transform}. Using its eigenbasis, we derive closed form expressions for the overlap matrix as a function of the phase space rotation, which can be used to compute the entanglement spectrum following the discussion of Section \ref{sec:ent}.

\subsection{Phase space rotation and the fractional Fourier transform}   \label{sec:rot_fracFT}
The phase space rotation acts on $\bxi = (x,p) \in \real^2$ as 
\beq 
  g_\theta \cn \bxi \mapsto \bxi' = 
  \left( \begin{array}{cc}
    \cos\theta & -\sin\theta \\ 
    \sin\theta & \cos\theta 
  \end{array} \right) \bxi.
\eeq 
Consider a position basis state $\ket{y}$, for which the Wigner function is 
\beq 
  \fl \qquad W(x,p) = \bra{y} \wigE(x,p) \ket{y} = \int \frac{\rmd x_1 \rmd p_1}{2\pi} \rme^{ p x_1 - i p_1 \left( x - \frac{1}{2} x_1 - y \right) } \delta(x_1) = \delta(x - y).
\eeq
Since the Wigner function transforms as a scalar, under a phase space rotation by $\theta$,
\beq 
  W(\bxi) \mapsto W_\theta(\bxi) \equiv W(g_\theta^{-1} \bxi) = \delta(x' \cos\theta + p' \sin\theta - y). 
\eeq
Using Equation (\ref{eq:wigner_inv}), the inverse WWT for this Wigner function can be computed as
\begin{eqnarray}
 \fl \qquad \hat\rho_\theta' & =  \int \rmd x_1 \rmd x_2 \left[ \int \frac{\rmd p'}{2\pi}  \,  \delta \left(\frac{x_1 + x_2}{2} \cos\theta + p' \sin\theta - y \right)  \,  \rme^{i p (x_1 - x_2)} \right] \ket{x_1}  \bra{x_2}  \nonumber \\ 
 \fl \qquad & =  \frac{1}{2\pi|\sin\theta|} \int \rmd x_1 \rmd x_2  \,\exp\left\{ i \left(y \csc\theta - \frac{x_1 + x_2}{2} \cot\theta \right) (x_1 - x_2)  \right\} \, \ket{x_1}  \bra{x_2}  \nonumber \\ 
 \fl \qquad & =  \frac{1}{2\pi|\sin\theta|} \int \rmd x_1 \, \rme^{ - \frac{i}{2} \cot\theta (x_1^2 - 2 x_1 y \sec\theta + y^2) }   \ket{x_1} \int \rmd x_2 \bra{x_2} \rme^{  \frac{i}{2} \cot\theta (x_2^2 - 2  x_2 y \sec\theta + y^2) }.
\end{eqnarray}
Thus, under a phase space rotation by $\theta$, the basis states transform as
\beq 
  \fl \qquad \ket{y}  \mapsto \frac{\rme^{i\phi(\theta)}}{\sqrt{2\pi|\sin\theta|}} \int \rmd x \, \rme^{ - \frac{i}{2} \cot\theta (x^2 - 2 x y \sec\theta + y^2) } \ket{x} 
  \equiv \int \rmd x \,  \uop_\theta(y,x) \ket{x}, 
\eeq 
so that an arbitrary wavefunction $\ket{\psi}$ transforms as 
\beq 
  \fl \qquad \ket{\psi} = \int \rmd x \, \psi(x) \ket{x} \to \int \rmd x \, \psi_\theta(x) \ket{x}, \qquad 
  \psi_\theta(x) = \int \rmd y \; \uop_\theta(x,y) \psi(y),   \label{eq:fracFT_def}
\eeq
which is uniquely defined up to an overall phase $\phi(\theta)$, owing to the projective nature of the representation. This representation can actually be made unitary by a suitable choice of $\phi(\theta)$. In \ref{app:fracFT}, we show that 
\[
  \uop_\theta \circ \uop_{\theta'} = \uop_{\theta + \theta'}   \implies \phi(\theta) + \phi(\theta') = \phi(\theta+\theta') + \frac{\pi}{4}.
\]
Setting $\phi(\theta) = \frac{\pi}{4} - \frac{\theta}{2}$, we obtain a unitary transformation for all $\theta$, which is the \emph{fractional Fourier transform}\cite{ozaktas_fracFT_book, ozaktas_fracFT, shen_fracFT, namias_fracFT},  well known to electrical engineers. The transformation can alternatively be expressed as 
\beq 
  \mathcal{F}_\theta \left[ \psi(x) \right] = \sqrt{ \frac{1-i\cot\theta}{2\pi} }  \int \rmd y \, \rme^{ - \frac{i}{2} \cot\theta (x^2 - 2 x y \sec\theta + y^2) } \psi(y).     
\eeq 
The transformation kernel reduces to the Fourier transform  for $\theta = \pi/2$, as well as to a Dirac-delta distribution as $\theta\to 0$, as shown in \ref{app:fracFT}. Finally, for $\theta\to\pi$, $\uop_\theta$ reduces to an inversion, i.e, $\psi(x) \mapsto -\psi(x)$. Thus, we have an explicit form for a continuous family of unitary operators on $\hilbert_1$ that interpolate between the identity and inversion operators, as alluded to in Section \ref{sec:phsp_cm}.

\subsection{Computing the overlaps} 
Given a set of single-particle wavefunctions $\ket{\psi_a}, \, a = 1, \dots N$, we can now compute the overlap matrix as a function of the phase space rotation angle as 
\beq 
  \overlap_{ab}(\theta) = \int_{x \in \sA} \rmd x \, \psi_{a,\theta}^\ast(x) \, \psi_{b,\theta}(x), \qquad 
  \psi_{a,\theta}(x) = \mathcal{F}_\theta [\psi_a(x)].
\eeq
However, since evaluating the integral in Equation \ref{eq:fracFT_def} in a closed form can in general be daunting, we take an alternative route. Recall that the fractional Fourier transform is a linear unitary operator on $L^2(\real)$, so that it has a complete set of eigenvectors with the eigenvalues on the unit circle. These are the eigenstates of the 1D quantum harmonic oscillator\cite{ozaktas_fracFT_book, namias_fracFT}, described by the Hamiltonian 
\beq 
  \hlt_{\text{SHO}} = \frac{1}{2} \left( \hat{p}^2 + \hat{x}^2 \right) = \ad \a + \frac{1}{2}, 
\eeq 
where the ladder operators are defined as 
\beq 
  \a = \frac{1}{\sqrt{2}} \left( \hat{x} + i \hat{p} \right), \qquad 
  \ad = \frac{1}{\sqrt{2}} \left( \hat{x} - i \hat{p} \right),    \label{eq:ladder_def}
\eeq
which satisfy $[\a, \ad] = 1$. The eigenvectors satisfy  
\beq 
  \hlt_{\text{HO}} \ket{\varphi_n} = \left( n + \frac{1}{2} \right) \ket{\varphi_n}, \qquad  \varphi_n(x) = \frac{1}{\sqrt{2^n \, n!}} H_n(x) \rme^{-x^2/2},
\eeq
where $H_n(x)$ are the Hermite polynomials. Under phase space rotations, these transform as 
\beq   
  \uop_\theta \ket{\varphi_n} = \rme^{i n \theta} \ket{\varphi_n}
  \iff \mathcal{F}_\theta[\varphi_n(x)] = \rme^{i n \theta} \varphi_n(x).
\eeq
This is simply the quantum mechanical time evolution of the harmonic oscillator eigenstates, with $\theta$ playing the role of ``time''! We can therefore rewrite $\uop_\theta = \rme^{i \, \ad \a}$, which can be thought of as the \emph{quantum version} of the classical statement (see Equation \ref{eq:rot_hlt_def}) that the harmonic oscillator Hamiltonian generates the rotation subgroup of $\ISp(2d, \real)$. Finally, the fractional Fourier transform kernel can also be interpreted as the propagator of the harmonic oscillator (\emph{Mehler's kernel}).

The phase space rotation can now be implemented by first expanding them in the harmonic oscillator eigenbasis, as
\beq 
  \ket{\psi_a} = \sum_{n=0}^\infty \alpha_{a,n} \ket{\varphi_n} \implies 
  \ket{\psi_{a,\theta}} = \sum_{n=0}^\infty \alpha_{a,n} \rme^{in\theta} \ket{\varphi_n}.    \label{eq:wf_HO_expn}
\eeq 
The half space overlaps (Equation \ref{eq:overlap_def}) are then given by 
\beq 
  \overlap_{ab}(\theta) = \braket{\psi_{a,\theta}}{\psi_{b,\theta}}_\sA = \sum_{m,n} \alpha_{a,m}^\ast \alpha_{b,n}^\pdg \, \rme^{i(n-m)\theta} \, \overlapHO_{mn},   \label{eq:overlap_psi}
\eeq 
where $\theta$-independent $\overlapHO_{mn} = \braket{\varphi_m}{\varphi_n}_\sA$ can be calculated explicitly, as shown in the next section. The overlap matrix can be more compactly written as 
\beq 
  \overlap(\theta) = \fA^\dg \Theta^\dg(\theta) \, \overlapHO \, \Theta(\theta) \fA, \qquad 
  \Theta = \diag{1, \rme^{i\theta}, \dots },     \label{eq:overlap_mat}
\eeq 
where $\fA = \left( \bal_1, \bal_2, \dots \bal_N \right)^T$. In practice, we truncate the expansion in Equation \ref{eq:wf_HO_expn} and use this expression to numerically evaluate the overlap matrix, and hence the entanglement spectrum, as a function of $\theta$. In the following, we term the entanglement spectrum as a function of $\theta$ the \emph{phase space entanglement spectrum}(PSES).

\subsection{Overlap matrix and Wigner functions}
\begin{figure}[ht]
  \centering
  \includegraphics[width = 0.9\textwidth]{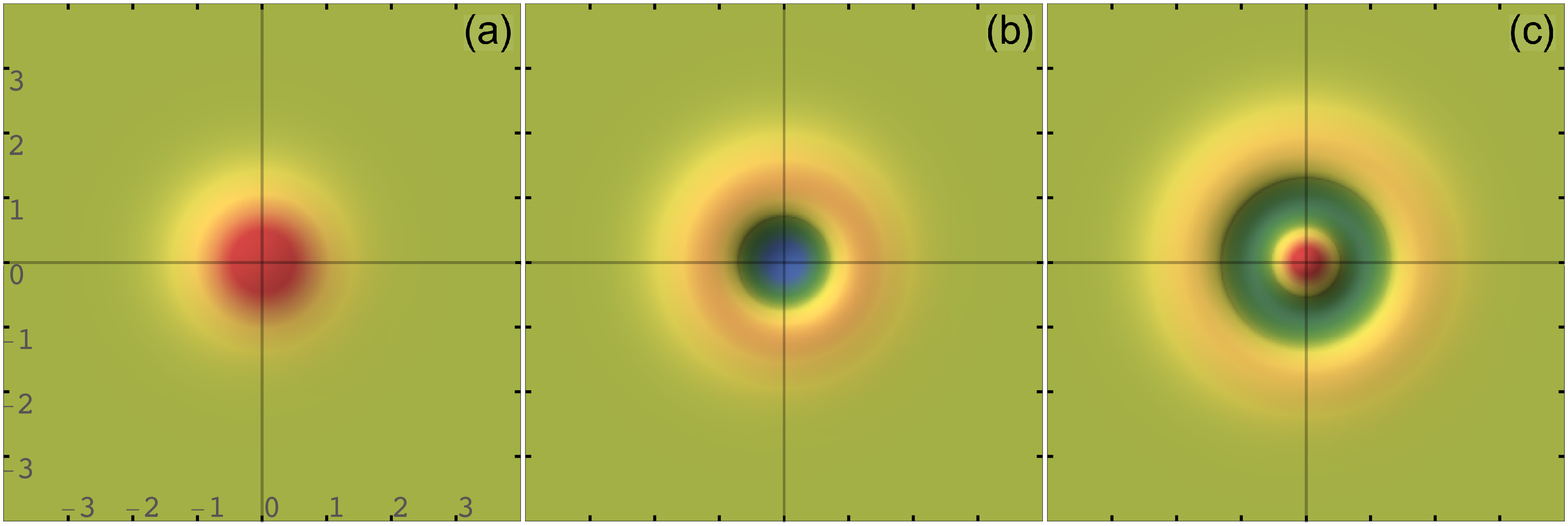}
  \includegraphics[width=0.062\textwidth]{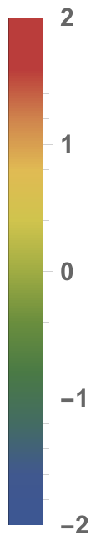} 
  \caption{The Wigner functions for the first three eigenstates of the harmonic oscillator. The two axes are $x$ and $p$, respectively. } 
  \label{fig:wignerHO}
\end{figure}

The half-space overlap matrices for the harmonic oscillator can be computed analytically using the harmonic oscillator Wigner functions. These are also useful in their own right, since given an arbitrary wavefunction $\ket{\psi} \in \hilbert_1$ with harmonic oscillator coefficients $\bal$, the Wigner function can be computed as
\beq
  \fl \qquad
  W(\bxi) = \sum_{m,n = 0}^\infty \alpha_m^\ast \alpha_n \mathrm{e}^{i(n-m)\theta} W_{mn}(\bxi); \qquad 
  W_{mn} (\bxi)  = \bra{\varphi_m} \wigE(\bxi) \ket{\varphi_n}.  
\eeq
These can then be used to visualize the single particle density  matrices for arbitrary wavefunctions. To compute $W_{mn}(\bxi)$, we use the harmonic oscillator creation and annihilation operators defined in Equation \ref{eq:ladder_def}. Mirroring these, we also define complex coordinates on the phase space as\footnote{ 
  We define these with the extra factor of $1/\sqrt{2}$ to get rid of the additional factor of 2 in the symplectic form, and consequently in the WWT kernel. 
} 
\beq 
  \coord = \frac{1}{\sqrt{2}} \left( x + i p \right), \quad 
  \coord^\ast = \frac{1}{\sqrt{2}} \left(x - i p \right),   \label{eq:coord_def}
\eeq
so that the symplectic form is $\omega = - i \rmd\coord \wedge \rmd \coord^\ast$. In these coordinates, the operator $\wige(\bxi)$ can be written as 
\begin{eqnarray}
  \wige (\coord) & =  \rme^{ -i \omega\left( \coord, \hat{\coord} \right) } 
  =  \rme^{ \coord^\ast \a - \coord \ad  } 
  = \rme^{-|\coord|^2/2} \rme^{-\coord \ad} \rme^{\coord^\ast \a },
\end{eqnarray}
and the Wigner functions become 
\beq 
  \fl \qquad W_{mn}(\coord) = \bra{\varphi_m} \wigE(\coord) \ket{\varphi_n} 
  = \frac{1}{2\pi} \int \rmd^2\coord_1 \; \rme^{\coord^\ast \coord_1 - \coord_1^\ast \coord} \;\rme^{-|\coord_1|^2/2} \bra{\varphi_m} \rme^{-\coord_1 \ad} \rme^{\coord_1^\ast \a} \ket{\varphi_n},
\eeq 
where the expectation value occuring in this integral can be computed using operator manipulations, as shown in \ref{app:HO_wigner}. The final result is (see also Refs \cite{justin_wigner_func, ripamonti_wigner_HO})
\beq  
  W_{mn} (\coord ) = 2 (-1)^m \sqrt{\frac{m!}{n!}} (2\coord )^{n-m} \rme^{-2 |\coord |^2} L^{n-m}_m (4|\coord |^2),
\eeq
where $L_n^\alpha(x)$ denote the associated Laguerre functions. Clearly, for $m = n$, we get
\beq 
  \fl\qquad W_{nn} (\coord) = 2 (-1)^n  \rme^{-2\abs{\coord}^2} L_n (4\abs{\coord}^2) = 2 (-1)^n  \rme^{-(x^2 + p^2)} L_n (2(x^2 + p^2)).    \label{eq:wigner_HO}
\eeq 
For pure harmonic oscillator eigenstates, the Wigner functions are circularly symmetric, and thus manifestly invariant under a phase space rotation. This is consistent with the projective nature of the WWT, since the wavefunctions do indeed change by a phase under phase space rotation. We plot the first three Wigner functions\footnote{
  These Wigner functions look identical to the quantum Hall wavefunctions in the lowest Landau level in the symmetric gauge. This is not surprising, since in the lowest Landau level, the two 	coordinates are canonically conjugate, thereby mimicking the noncommutative nature of the phase space. 
}
in Figure~\ref{fig:wignerHO}.

Finally, we compute the half-space overlap matrix $\overlapHO_{mn} = \braket{\varphi_m}{\varphi_n}_\sA$. Note that $\varphi_n(-x) = (-1)^n \varphi_n(x)$, so that from orthonormality, we get 
\beq 
  \delta_{mn} = \int_{-\infty}^\infty \rmd x \, \varphi_m^\ast(x) \varphi_n^\pdg(x) = \left[ 1 + (-1)^{m+n} \right] \int_0^\infty \rmd x \, \varphi_m^\ast(x) \varphi_n^\pdg(x),    \label{eq:HO_overlap_even}
\eeq 
which fixes the overlap if $m+n$ is even. For $m+n$ odd, we use the explicit form of the cross Wigner function $W_{mn}(\bxi)$ and integral in Equation \ref{eq:wig_overlap} to compute $\overlapHO_{mn}$, as shown in \ref{app:HO_wigner}. The final result is 
\begin{eqnarray} 
 \overlapHO_{mn} & = \int_0^\infty \rmd x \, \varphi_m^\ast(x) \varphi_n^\pdg(x) = 
 \cases{
    \delta_{mn}/2, & $m+n$ even \\   
    \mathfrak{o}_{mn}, & $m + n$ odd },  \label{eq:overlapHO}
\end{eqnarray}
where 
\[
 \fl \quad \mathfrak{o}_{mn} = \frac{i^{m-n+1}}{2\pi} \sqrt{\frac{n!}{m!}}  \frac{(-1)^m 2^{(n-m)/2} }{(n-m)!}  \Gamma\left( \frac{n-m}{2}  \right)  \hypgeo \left( -m, \frac{n-m}{2} + 1, n-m+1; 2 \right),
\]
and $\hypgeo(a,b,c; x)$ denotes the ordinary hypergeometric function\cite{bateman_spcl}.

\section{Examples: Inversion-symmetric wavefunctions}   \label{sec:res}
\newcommand{\Mmat}{\mathfrak{M}}
\newcommand{\omat}{\mathfrak{m}}
\newcommand{\oeig}{\mathfrak{m}}
\newcommand{\onum}{\mathfrak{n}}

In this section, we illustrate the analytical machinery derived in the last three sections by computing the entanglement spectra as a function of phase space rotation for inversion symmetric free fermion many-body states. More explicitly, given the orthonormalized set of single particle wavefunctions that constitute the Slater determinant, we use Equations \ref{eq:overlap_mat} and \ref{eq:overlapHO} to compute the half-space overlap matrices as a function of $\theta$, and thus compute the entanglement spectrum using Equation \ref{eq:redc_dens_mat}. 

The entanglement spectra of inversion symmetric Slater determinants are of particular interest, since they are known to exhibit a ``chiral symmetry'' for the position space cut\cite{ashvin_ent_inv, ryu2006entanglement}. In this section, we show that this feature survives phase space rotations by arbitrary $\theta$ about the inversion center. For an even number of particles, one can further define an $\intg$-valued ``chiral invariant'' associated with the $\theta$-dependent entanglement spectrum, thereby classifying the entanglement Hamiltonians into topological sectors, which cannot be continuously deformed into each other without closing the gap at some $\theta$.

\subsection{Inversion symmetry and phase space rotation}
For a many-body state $\ket{\Psi}$, inversion symmetry is the statement that 
\begin{eqnarray}
  &\Psi(-\mathbf{X}) = (-1)^\sigma\Psi(\mathbf{X}), \qquad \mathbf{X} = (x_1,\ldots, x_N) \in \mathbb{R}^{N}, 
 \end{eqnarray}
where $\sigma \in \{0,1\} $ is the mirror parity of the $N$-fermion position space wavefunction $\Psi(\mathbf{X}) = \langle \mathbf{X}|\Psi\rangle$.  For a position space cut about $x=0$, the entanglement spectrum is symmetric about $\ve_\sE = 0$. This can be seen from the results of Section \ref{sec:ent}, since inversion swaps the A and B subsystems, so that the overlap eigenvalues change as $\eig_a \to 1-\eig_a$ and using  Equation \ref{eq:ent_eng_def}, we get $\ve_{E,a} \to -\ve_{E,a}$. In the following, we show that this chiral symmetry of the entanglement spectrum stays intact under a phase space rotation.

For an inversion symmetric Slater determinant, we can always choose the corresponding orthonormal single particle states as inversion eigenstates by a suitable $U(N)$ rotation. In particular, given the single particle states $\ket{\phi_n}, n = 1, \dots N$, this can be achieved by diagonalizing the inversion operator $\mathcal{I}_x$ on $\text{span}\{\ket{\phi_1}, \ldots, \ket{\phi_N}\}$, i,e, by diagonalizing the matrix 
\begin{equation}
  (\mathcal{I}_x)_{mn} = \bra{\phi_m} \mathcal{I}_x \ket{\phi_n} = \int_\mathbb{R} \rmd x \; \phi_m^\ast(x) \phi_n(-x).    
\end{equation}
This matrix has eigenvalues $\pm 1$, and its eigenvectors can be used to define the single particle wavefunctions $\ket{\psi_n}$, which satisfy 
\begin{equation}
  \psi_n(-x) = (-1)^{\sigma_n}\psi_n(x); \qquad \sigma = \sum_{n=1}^N \sigma_n \; \text{mod}\;\; 2 = N_\rmo \; \text{mod}\;\; 2,
\end{equation}
where $\sigma_n \in \{0,1\}$ and $N_\mathrm{e,o}$ denotes the number of single particle wavefunctions with even/odd parity under inversion, such that $N=N_\rme+N_\rmo$. Using the orthogonality of these wavefunctions(see Equation \ref{eq:HO_overlap_even}), we deduce that 
\beq 
  \delta_{mn} = \braket{\psi_m}{\psi_n} = \left[ 1 + (-1)^{\sigma_m + \sigma_n} \right]\overlap_{mn}   \label{eq:Omn}
\eeq 
so that $\overlap_{mn} = \delta_{mn}/2$ whenever $\sigma_m = \sigma_n$, i.e, $\overlap$, restricted to a fixed parity sector, is proportional to identity.

Remarkably, this form of $\overlap$ continue to hold for phase space rotations by $\theta$ about $x=0$, the fixed point of inversion. This can be explicitly seen by switching to the harmonic oscillator basis, where phase space rotations take a simple form: 
\beq 
  \fl \qquad
  \ket{\psi_n} = \sum_{\ell=0}^\infty \gamma_{n,2\ell+\sigma_n} \ket{\varphi_{2\ell+\sigma_n}} \implies 
  \ket{\psi_{n,\theta}} = \sum_{\ell=0}^\infty \rme^{i(2\ell+\sigma_n)\theta} \gamma_{n,2\ell+\sigma_n} \ket{\varphi_{2\ell+\sigma_n}}.  
\eeq 
In the expansion, we have used the fact that under inversion, $\ket{\varphi_n} \to (-1)^n \ket{\varphi_n}$.
Orthonormalization again demands that $\sum_\ell \gamma_{m,\ell}^\ast \gamma^\pdg_{n,\ell} = \delta_{mn}$. so that if $\sigma_m = \sigma_n$, then
\begin{eqnarray}
  \fl \qquad \overlap_{mn}(\theta) 
  &= \sum_{\ell,\ell'} \gamma_{m,2\ell+\sigma_m}^\ast \gamma_{n,2\ell'+\sigma_n}^\pdg \rme^{2i(\ell'-\ell)\theta} \braket{\varphi_{2\ell+\sigma_m}}{\varphi_{2\ell'+\sigma_n}}_\sA \nonumber \\
  &= \frac{1}{2}  \sum_\ell \gamma_{m,2\ell+\sigma_m}^\ast \gamma_{n,2\ell'+\sigma_n}^\pdg  = \frac{1}{2} \delta_{mn},
\end{eqnarray}
independent of $\theta$, where we have used Equation \ref{eq:overlapHO} for the harmonic oscillator eigenstates. This result can also be understood by the fact that a phase space rotation cannot change the parity sector of a wavefunction, since inversion, being equivalent to $\uop_\pi$, commutes with phase space rotation $\uop_\theta$.

\subsection{Entanglement spectrum and the chiral invariant} 
We next investigate how the additional constraints imposed on the overlap matrix affect the entanglement spectrum. The structure of the overlap matrix is made particularly clear by enumerating the wavefunctions such that $\sigma_m = 0$ for $1 \leq m \leq N_\rme$ and $\sigma_n = 1$ otherwise. Then, $\overlap(\theta)$ can be written as
\beq
  \overlap(\theta) = \frac{1}{2} \left[ \id + \Mmat(\theta) \right], \qquad 
  \Mmat(\theta) = 2\left( \begin{array}{cc} 
          0 & \omat(\theta) \\ 
          \omat^\dg(\theta) & 0
         \end{array}
 \right), \label{eq:overlap_block}
\eeq
with 
\beq 
  \omat_{mn}(\theta) = \braket{\psi_{m,\theta}}{\psi_{N_\rme + n,\theta}}_A, \qquad m = 1, \dots N_\rme, \, n = 1, \dots N_\rmo. 
\eeq
Using Equation \ref{eq:ent_hlt}, the entanglement Hamiltonian becomes 
\beq 
   \hlt_\sE(\theta) = -\ln \left[ \left( \id + \Mmat(\theta) \right) \left( \id - \Mmat(\theta) \right)^{-1} \right] = -2 \tanh^{-1} \Mmat(\theta).  
\eeq
Thus, $\hlt_\sE(\theta)$ inherits the off-diagonal structure of $\overlap(\theta)$, and hence enjoys a chiral symmetry. Explicitly, 
\beq 
    \{ \Mmat(\theta), \mathcal{I}_x \} =\{ \hlt_\sE(\theta),\mathcal{I}_x \} = 0, \qquad 
\mathcal{I}_x = 
\left( \begin{array}{cc} 
          \id_{N_\rme} & 0 \\ 
          0 & -\id_{N_\rmo}
         \end{array}
 \right).
\eeq
Unlike the chiral symmetry typically encountered in topological band theory, the origin of this entanglement chiral symmetry is simply the invariance of the structure of the overlap matrix  under phase-space rotations.

For a Slater determinant with $N_\rme \neq N_\rmo$, the entanglement Hamiltonian $\hlt_\sE(\theta)$ has at least $|N_\rme-N_\rmo|$ zero modes for all $\theta$, since $\rank \hlt_\sE = \rank \Mmat \leq 2 \min(N_\rme, N_\rmo)$. These zero energy `flat bands' in the 1-particle entanglement spectrum are protected by the inversion symmetry, and lead to a lower bound on the entanglement entropy\cite{ryu2006entanglement} 
\beq 
  S_\sA \geqslant  \ln 2 \, |N_\rme-N_\rmo|.  
\eeq
This lower bound can be associated with the structure of the boundary modes in the context of topological band theory\cite{ryu2006entanglement}.

On the other hand, for $N_\rme = N_\rmo$, the entanglement Hamiltonian is generically gapped for all $\theta$, and thus resembles Bloch Hamiltonians of a chiral symmetric 1D topological band insulator! Thus, we can exploit the topological classification\cite{chiu2016classification} of band insulators with symmetries to define a \emph{chiral invariant}. Explicitly, the gap in the entanglement spectrum closes iff $\omat(\theta)$ is singular, so that for gapped Hamiltonians, we can define the chiral invariant as 
\beq 
  \nu_\sE \equiv \frac{1}{\pi} \text{Im} \left[ \int_0^\pi    \Tr [\omat^{-1} \! (\theta) \, \partial_\theta \omat(\theta)] \, \mathrm{d}\theta \right].    \label{eq:chiral_inv}
\eeq
The range of integration here reflects the fact that $\uop_\pi = \mathcal{I}$, which returns $\ket{\Psi}$ to itself up to the sign $(-1)^{N_\rmo}$. Mathematically, the topological information associated with $\omat(\theta)$ is reflected in the fact that $\omat \cn S^1 \to \GL(N_\rme, \cmplx)$, and such maps are classified by $\pi_1 \left( \GL(N_\rme, \cmplx) \right) \cong \intg$. Equivalently, the chiral invariant is the winding number associated with the map $\theta \mapsto \det \omat(\theta)$. 

The existence of this topological invariant is surprising, since it is well defined for Slater determinants with any number of fermions; the simplest case being that of two fermions. It can also be used to define various topological `phases', which cannot be deformed into each other without closing the gap in the entanglement spectrum, corresponding to a topological `phase transitions' between different quantized values of $\nu_\sE$. These phase transitions are again reflected in a lower bound in the entanglement entropy: for $\mathscr{N}_\text{zeros}$ Dirac-like crossings at critical angles $\theta$, we must have 
\beq
S_\sA \geqslant 2 \ln 2 \, \mathscr{N}_\text{zeros}. 
\label{eq:S_bound}
\eeq
One could interpret this lower bound as the minimum amount quantum entanglement that needs to be introduced into an intermediate pure state during the course of smoothly and unitarily transforming between inversion symmetric Slater wavefunctions with different $\nu_\sE$ index.  


\subsection{Flat bands and topological phase transitions}
We now consider a simple example of a Slater determinant for which the computation of the winding number is analytically tractable. Consider then the $N$-fermion states formed by the eigenstates of the simple harmonic oscillator, i.e, 
\beq 
  \ket{\Psi} \equiv \ket{\varphi_{m_1}} \wedge  \dots \wedge  \ket{\varphi_{m_N}}, \qquad m_j \in \intg^+, \; 0 \leq m_1 < m_2 < \dots m_N.
\eeq 
Under a phase space rotation, $\ket{\varphi_n} \to e^{in\theta} \ket{\varphi_n}$, so that the Slater determinant changes only up to a global phase $\ket{\Psi} \mapsto e^{i \theta \sum_j m_j} \ket{\Psi}$, i.e, the many-body density matrix, and hence the reduced density matrix, is invariant under phase space rotations. Thus, the entanglement spectrum is independent of $\theta$, leading to a ``flat band'' model, which may nonetheless carries a nontrivial chiral invariant when $N_\rmo = N_\rme$. 

Setting $N = 2M$ and given a $\vm = (m_1, \dots m_{2M})$ for which the chiral invariant is defined, it can be computed by referring back to Equations~\ref{eq:overlap_mat}, and noting that $\bal$ reduces to an identity matrix under row and column operations, while separating out matrices into the parity-odd and even sectors such that $\Theta = \text{diag} \{ \Theta_\rme, \Theta_\rmo \}$, with $\Theta_{\rmo/\rme}$ being the restriction of $\Theta$ defined in Equation~\ref{eq:overlap_mat} to the relevant odd/even sector. The overlap matrix takes the form 
\beq 
  \Theta^\dg \overlapHO \Theta = 
  \Theta = \frac{1}{2} \left(
  \begin{array}{cc}
    \id & 2 \Theta_\rme^\dg \wt{\overlapHO} \Theta_\rmo  \\ 
    \Theta_\rmo^\dg \wt{\overlapHO} \Theta_\rme  & \id
  \end{array}
  \right) \implies 
  \omat(\theta) = \Theta_\rme^\dg \wt{\overlapHO} \Theta_\rmo,
\eeq 
where $\wt{\overlapHO}_{mn} = \overlapHO_{m+M,n}$. Thus, the expression for the chiral invariant in Equation~\ref{eq:chiral_inv} reduces to 
\beq 
  \nu_\sE = \frac{1}{\pi} \text{Im} \left[ \int_0^\pi \Tr [\Theta_\rmo^\dg \partial_\theta \Theta_\rmo + \Theta_\rme \partial_\theta \Theta_\rme^\dg ] \, \mathrm{d}\theta \right],
\eeq
which can be evaluated explicitly, since $\Theta_\rmo$ and $\Theta_\rme$ are diagonal matrices. For instance, consider the $2M$-fermion ground state, for which $m_j = j-1$, so that the odd/even sectors of $\Theta$ are simply $\Theta_\rmo = \rme^{i\theta} \Theta_\rme = \text{diag}\{ \rme^{i\theta}, \rme^{i 3\theta},  \dots \rme^{i(2M-1)\theta} \}$. The chiral invariant becomes 
\beq 
  \nu_\sE = \sum_{n=0}^{M-1} \left( 2n+1 \right) - \sum_{n=0}^{M-1} 2n = M.   \label{eq:chiral_inv_GS}
\eeq 
On the other hand, for the first excited state with $m_j = j$, the chiral invariant becomes 
\beq 
  \nu_\sE = \sum_{n=0}^{M-1} \left( 2n+1 \right) - \sum_{n=1}^{M} 2n = -M.     \label{eq:chiral_inv_ES}
\eeq 
Thus, a flat PSES with any nonzero chiral invariant can be realized by a many-fermion ground- or first excited-state of the simple harmonic oscillator. Consequently, any many-body wavefunction for which the chiral invariant is well-defined can be continuously deformed into once of these harmonic oscillator many-fermion states without closing the entanglement gap. 

\begin{figure}
  \includegraphics[width = 0.93\textwidth]{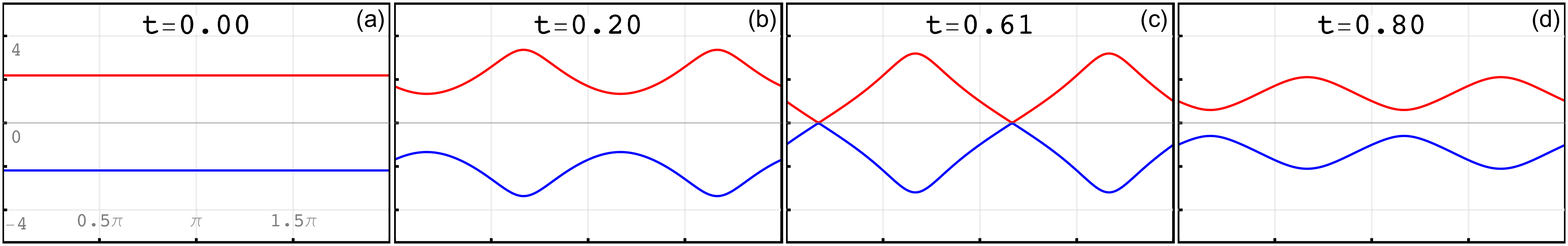} \\ \vspace{-0.18in} \\ 
  \includegraphics[width = 0.93\textwidth]{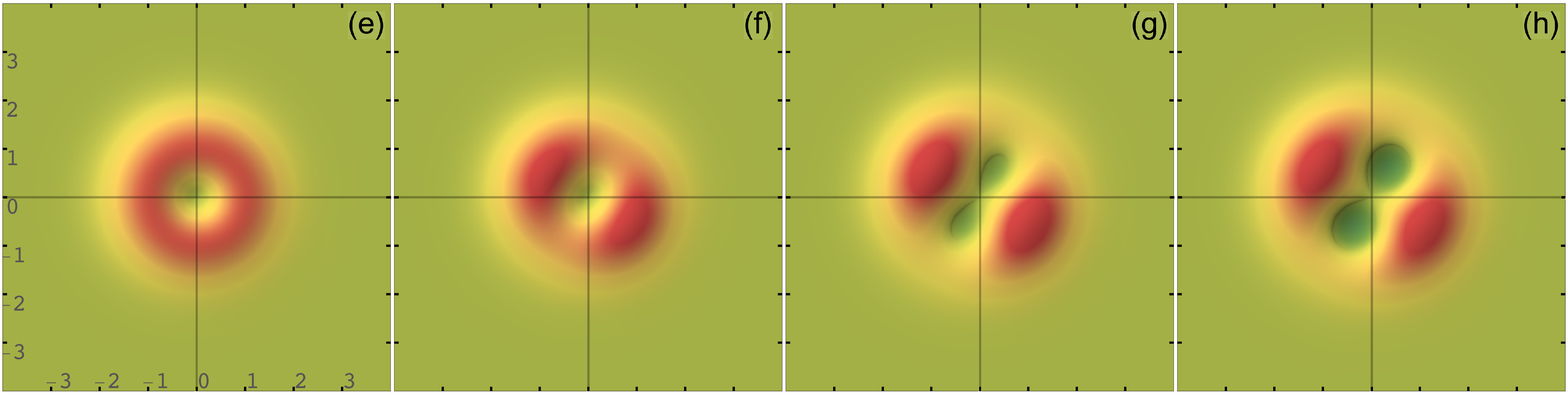}
  \includegraphics[width=0.046\textwidth]{legend.eps} 
  \caption{
    (Top row) The one particle entanglement spectrum as a function of phase space rotation angle $\theta \in (0,2\pi)$ for the 2-fermion excited states of the simple harmonic oscillator described in Equation \ref{eq:psi_t},  with $\phi=2\pi/3$ for various values of $t$. (Bottom row) The corresponding Wigner functions for the 1-particle density matrix. The winding number changes from  $\nu_\mathrm{E} = -1$ in (a) to $\nu_\mathrm{E}=1$ in (d), via a gap closing for $t =  \frac{2}{\pi} \tan^{-1} \sqrt{2} \approx 0.61$. 
  }
  \label{fig:ES_excited}
\end{figure}

\begin{figure}
  \begin{center}
  \includegraphics[width = 0.45\textwidth]{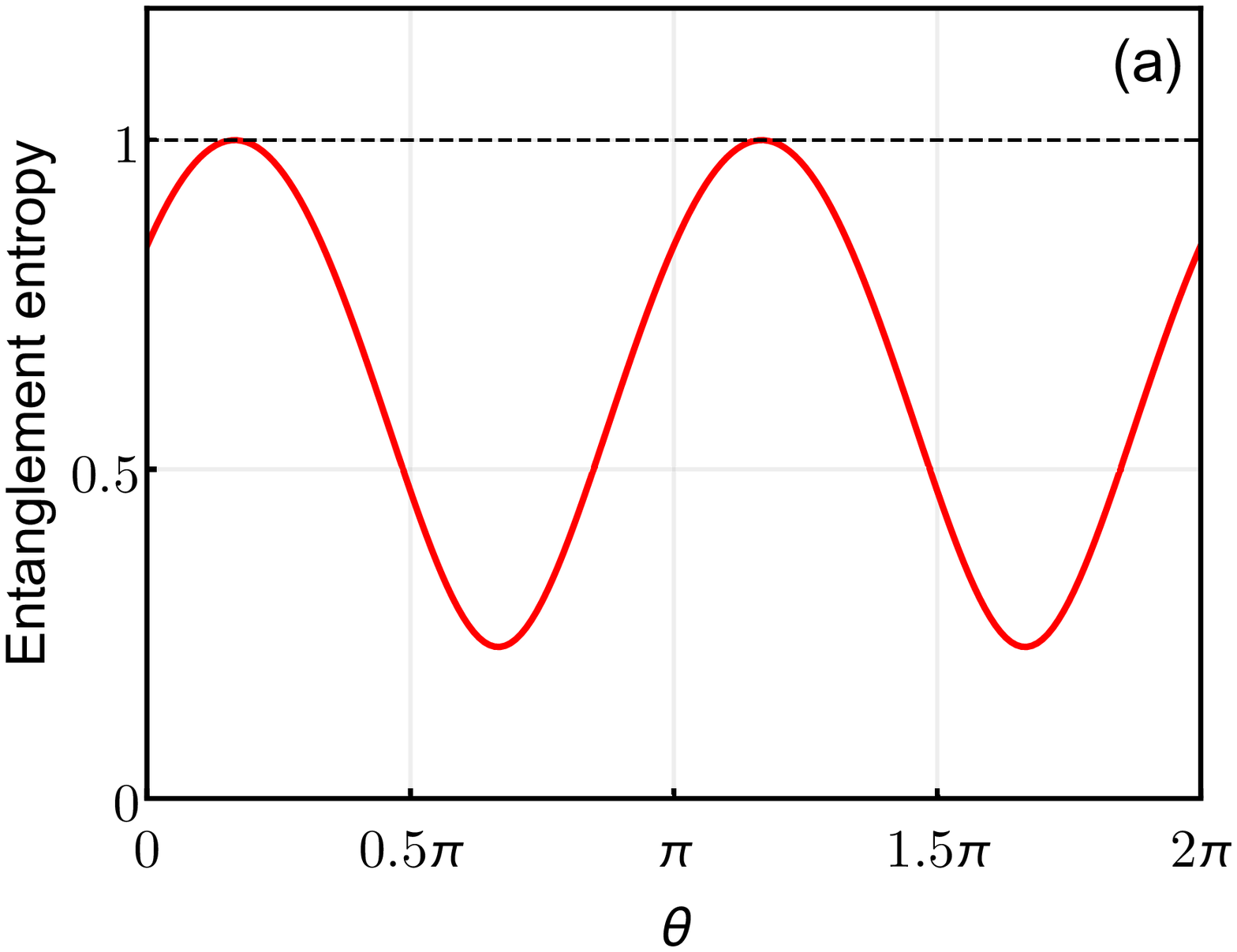}
  \includegraphics[width = 0.45\textwidth]{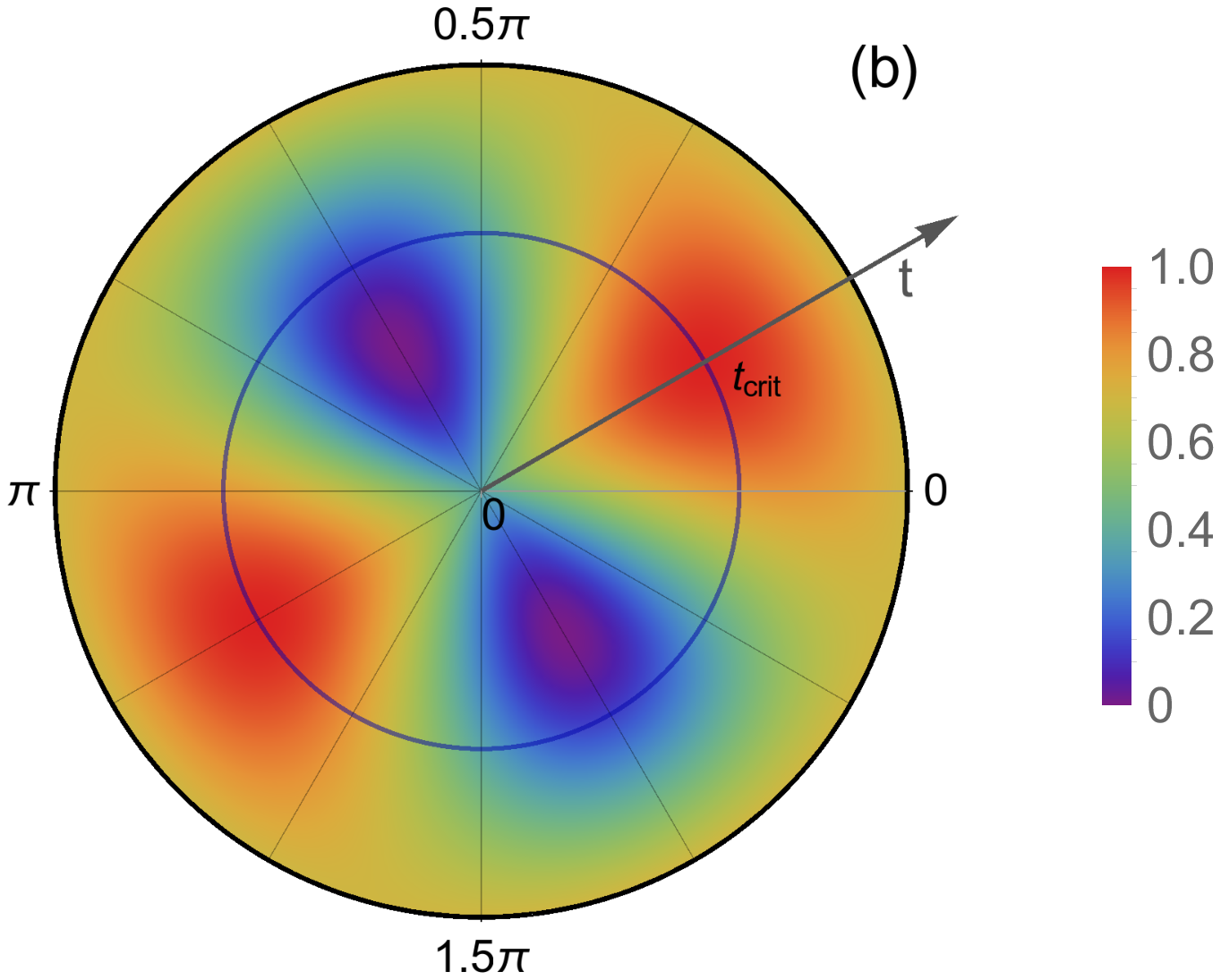}
  \end{center}
  \caption{
    (a) The entanglement entropy $S_A/2 ln(2)$ of the excited state described in Equation~\ref{eq:psi_t} for $t_\text{crit} \approx 0.6$ as a function of the phase space rotation angle $\theta\in [0,2\pi]$, which saturates the entanglement entropy bound of Equation~\ref{eq:S_bound}, since there are only 2 particles. (b) The entanglement entropy as a function of $t\in[0,1]$ as well as $\theta\in [0,2\pi]$ as a polar plot, which shows that the entanglement entropy is maximum for $t = t_c$.}
  \label{fig:SA_polar}
\end{figure}

The chiral invariant provides a topological classification of various gapped entanglement spectra. Thus, a deformation between wavefunctions with gapped PSES carrying different chiral invariants must proceed via the closing of the gap at some $\theta \in [0,\pi]$, which is an analogue of a \emph{topological phase transition} in the topological band theory. This gap closing can also be analytically investigated for the harmonic oscillator wavefunctions. Consider then the 2-fermion ground state(GS) and the first excited states(ES), with wavefunctions 
\begin{equation}
  \ket{\Psi_{\text{GS}}} := \ket{\varphi_0} \wedge \ket{\varphi_1}, \qquad  \ket{\Psi_{\text{ES}}} := \ket{\varphi_2} \wedge \ket{\varphi_1}. 
\end{equation}
These two states have the same total parity $\sigma=1$, but different winding numbers $\nu_\sE = +1$ and $-1$, respectively, as follows from Equations~\ref{eq:chiral_inv_GS} and \ref{eq:chiral_inv_ES}. Now consider a smooth interpolation between these two states, given by 
\begin{equation}
  \ket{\Psi(t)} := \cos\left(\frac{\pi t}{2} \right) \ket{\Psi_{\text{GS}}} + \mathrm{e}^{i \phi} \sin\left(\frac{\pi t}{2} \right) \ket{\Psi_{\text{ES}}}, \qquad t \in [0,1] 
  \label{eq:psi_t}
\end{equation}
with $\phi \in \mathbb{R}$ being an arbitrary phase. We can explicitly compute 
\beq 
  \omat(\theta) =  \overlapHO_{01} \cos\left(\frac{\pi t}{2} \right) \rme^{i\theta} + \overlapHO_{21} \sin\left(\frac{\pi t}{2} \right) \rme^{i(\phi-\theta)}, 
\eeq
which vanishes for some $\theta$ iff $t_\text{crit} = \frac{2}{\pi} \tan^{-1} \left( \overlapHO_{01}/\overlapHO_{21} \right) = \frac{2}{\pi} \tan^{-1} \sqrt{2} \approx 0.608$.

In Figure~\ref{fig:ES_excited}, we plot the PSES for various values of $t$. We clearly see the entanglement gap closing for $t = t_{\text{crit}}$ at $\theta = (\pi+\phi)/2$, which is accompanied by a \emph{pinch-off} at the origin in the single-particle Wigner function. In Figure~\ref{fig:SA_polar}, we plot the entanglement entropy $S_\sA$ as a function of $(\theta,t) \in [0,2\pi]\times[0,1]$, which indeed shows that the minimum bound in the inequality of Equation~\ref{eq:S_bound} is saturated at the gap-closing points.


\subsection{Inversion symmetric potentials}
We finally consider more general inversion symmetric Slater determinants. A plethora of such wavefunctions are provided by the many-body ground states of one-dimensional potential wells. Consider then a parity even potential well, i.e, $V(-x) = V(x)$, that supports bound states with energies $E_0 < E_1 < \dots$. The corresponding eigenvectors satisfy $\psi_n(-x) = (-1)^{n+1} \psi_n(x)$, as follows from the Sturm-Liouville oscillation theorem\cite{simon2005sturm}. Due to the alternating inversion parities for eigenstates, for the $N$-fermion state, we have $N_\rme = N_\rmo+1$ whenever $N$ is odd, and $N_\rme=N_\rmo$ whenever $N$ is even. 

\begin{figure}
  \includegraphics[width=\textwidth]{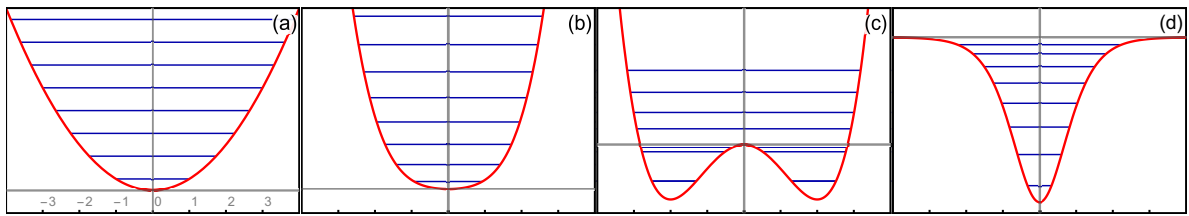}
  \caption{
    The form of potential wells (a) the simple harmonic oscillator, $V(x) = \frac{1}{2} x^2$, (b) the anharmonic oscillator $V(x) = \frac{1}{2} x^2 + \frac{1}{4} x^4$, (c) the double well $V(x) = -2x^2 + \frac{1}{4} x^4$ and the P\"oschel-Teller potential $V(x) = -45 \, \sech^2 x$, plotted over $x \in (-4,4)$. The energies of the first eight bound states are also depicted. 
  }
  \label{fig:potentials}
  
\end{figure}

\begin{figure}
  \includegraphics[width=\textwidth]{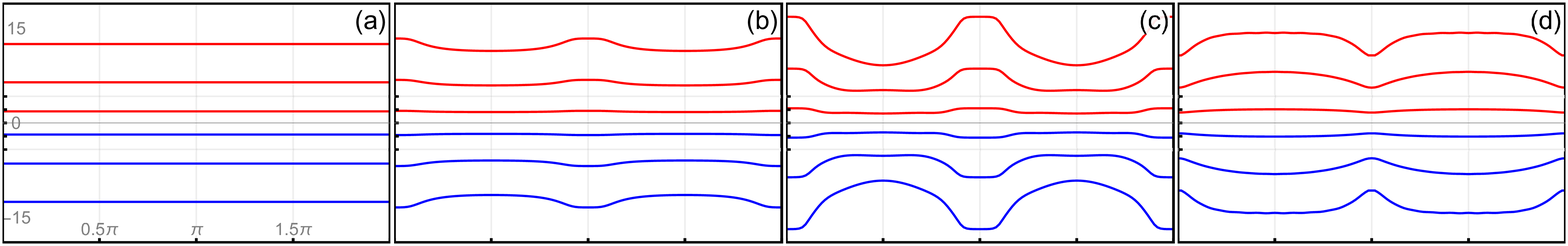}  \\ \vspace{-0.12in} \\ 
  \includegraphics[width=\textwidth]{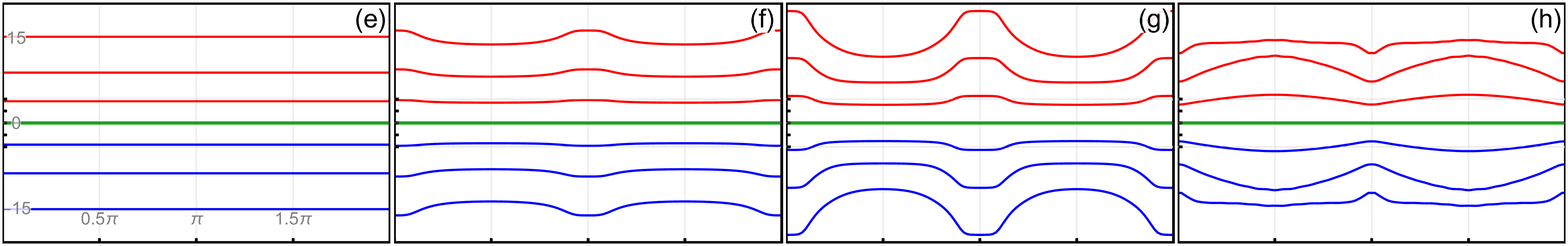} 
  \caption{
    The one particle entanglement spectrum as a function of phase space rotation angle $\theta \in (0,2\pi)$ for the N-fermion ground state of the potentials depicted in Figure~\ref{fig:potentials}, with $N = 6$ (top row) and $N=7$ (bottom row). In both cases, we clearly see the chiral symmetry of the spectrum. For odd $N$, there are flat zero energy bands (green lines), while for even $N$, there are associated winding numbers $\nu_\sE = 3$ in (a)-(d), respectively. 
    }
  \label{fig:ES_potentials}
\end{figure}

\begin{figure}
  \includegraphics[width=\textwidth]{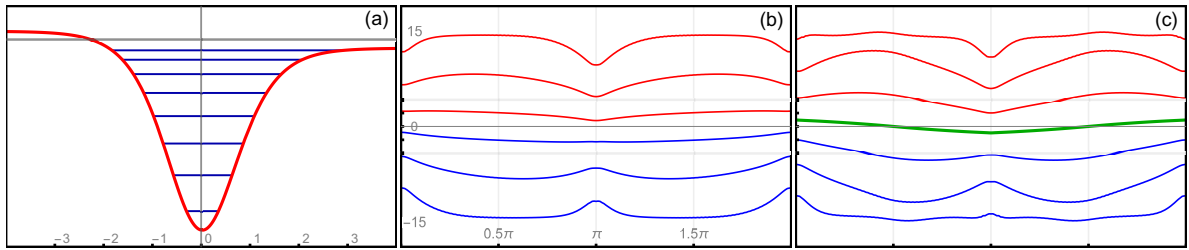}
  \caption{
    (a) The Rosen-Morse potential $V(x) =   -45 \, \sech^2 x - 2 \tanh x$, a variant of the P\"oschel-Teller potential which breaks the inversion symmetry. The corresponding N-fermion PSES for $N=6$ and $N=7$ particles are plotted in (b) and (c), respectively. which clearly show a breaking of the chiral symmetry. 
  }
  \label{fig:inv_breaking}
\end{figure}

The inversion symmetric potentials considered in this article, alongwith their bound state energies, are depicted in Figure~\ref{fig:potentials}. In order to compute their PSES, we rewrite the Hamiltonian in the harmonic oscillator basis and truncate at $M = 100$ basis states. A numerical diagonalization of the Hamiltonian then yields the harmonic oscillator coefficients $\bal$ for the first few bound states. The overlap matrix $\overlap(\theta)$ is then be computed using Equation~\ref{eq:overlap_mat}, which can then be used to compute the single particle PSES using Equation~\ref{eq:ent_hlt}. 

The results of these computations for both even and odd $N$ are plotted in Figure~\ref{fig:ES_potentials}. The simple harmonic oscillator exhibits flat band entanglement spectra, as discussed earlier. In the case of odd $N$, the presence of a zero entanglement energy flat band is observed, while in the case of even $N$, the entanglement spectrum is generically gapped with quantized $\nu_\mathrm{E}$ indices, which, for the potentials considered, depend only on $N$ and not on the potential in question. 

To demonstrate the role of inversion symmetry in the chiral symmetry of the phase space entanglement spectra, we consider an inversion asymmetric potential, viz, the Rosen-Morse potential (Figure~\ref{fig:inv_breaking}(a)), which is obtained by adding a parity odd term to the P\"oschel-Teller potential. The chiral symmetry is clearly broken in the corresponding PSES, as shown in Figure~\ref{fig:inv_breaking}.


\section{Discussion and conclusions}   \label{sec:conc}
In this article, we have generalized the conventionally studied entanglement cuts for fermionic many-body systems to a continuous family of cuts derived from the corresponding classical phase space, and introduce a general recipe for the computation of the entanglement spectra as a function of these cuts. We use these results to explicitly compute the entanglement spectra as a function of a phase space rotation, which continuously interpolate between the position- and momentum-space cuts. 

The phase space entanglement Hamiltonian possesses a chiral symmetry, which is a direct consequence of the invariance of inversion parity under phase space rotations. This then leads to a classification of all inversion symmetric free fermion wavefunctions in one spatial dimension, which belong to two broad classes depending on whether there are an unequal or equal number of even and odd parity modes. The former case always leads to flat zero energy entanglement bands, while the latter leads to gapped entanglement spectra that may be further classified by a topological winding number. This mirrors the classification of noninteracting topological phases of matter, which is intriguing, since it is well defined for as few as two fermions. 

We emperically notice that the topological phase transitions between winding states are accompanied by changes to the critical points of the Wigner function, as seen by the emerging of a saddle point in Figure~\ref{fig:ES_excited}. Since such a transition in a function on a compact manifold is often associated with a change in the Morse index, we speculate that the chiral invariant defined in this article should be related to the Morse index associated with the Wigner function, defined on the one-point compactified $\real^2$, i.e, on a 2-sphere. 

The phase space rotations studied in this aricle can be ``physically'' interpreted in various ways. In terms of geometric quantization, it can be thought of as starting with the prequantum Hilbert space and varying the polarization. In this picture, all possible entanglement cuts should be related by symplectomorphisms in the corresponding classical phase space, and one is simply choosing a different ``curve'' than the conventionally studied one in the group of symplectomorphisms. Interpreting the phase space rotation as the propagator of the quantum harmonic oscillator, i.e, interpreting $\theta$ as ``time'', the phase space entanglement spectrum can also be interpreted as a periodic time-dependent entanglement spectrum of, for instance, a wavepacked trapped in a harmonic potential well. 

The formalism and results discussed in this paper strictly apply only to continuum systems and finite number of single-component fermions; however, many aspects can be carried over to more general situations. The generalization to wavefunctions with internal degrees of freedom, a setup where entanglement is used to study noninteracting topological phases of matter, simply requires a redefinition of the overlap matrix using the suitable inner products. For discreet systems\cite{wootters1987wigner}, one would also need a discrete version of the fractional Fourier transform, which can be defined numerically as a fractional power of the (finite-dimensional, unitary) Fourier transform matrix\cite{ozaktas_fracFT_book}. Finally, a straightforward generalization to interacting fermionic wavefunctions may be performed by employing one particle density matrices obtained from marginalizing multi-determinant wavefunctions, in the spirit of Ref~\cite{schliemann2001quantum}. 

The study of entanglement has led to a deeper understanding of many aspects of quantum many-body systems. The phase space entanglement provides additional insight into the topology associated with certain inversion-symmetric systems, but further work is still needed to see if the combination of phase space based entanglement spectra and the analytical tools developed for the study of topological insulators can lead to new physical insights.




\ack
VD is funded by the Deutsche Forschungsgemeinschaft (DFG) with the CRC network TR 183 (Project B03). 
VC was supported by the Gordon and Betty Moore Foundation’s EPiQS Initiative through Grant GBMF4305 while at the University of Illinois. 
VD and VC gratefully acknowledge useful discussions with S. Ramamurthy and H Legg. VC is indebted to S. Low for exposing him to group theory on symplectic manifolds as a graduate student.

\appendix 

\section{Entanglement and fermionic many-body Hilbert space}    \label{app:hilbert}
The defining feature of fermionic wavefunctions is the antisymmetry under the exchange of particles. Mathematically, this can be encoded in an antisymmetric tensor product. Given a set of Hilbert space $V_n$, define
\beq 
  \bigwedge_{n=1}^N V_n \equiv \mathcal{A} \left[ \bigoplus_{\sigma\in S_N} V_{\sigma(1)} \otimes V_{\sigma(2)} \otimes \dots \otimes V_{\sigma(n)}  \right],
\eeq 
where $\mathcal{A}[\, ,]$ denotes the projection to the totally antisymmetric sector under the permutation group  $S_N$. More explicitly, if $V_n = \text{span}\{\uvec^n_1, \dots \uvec^n_{N_n}\}$, then 
\[
  \fl \qquad V_1 \wedge V_2 = \mathcal{A}\left[ (V_1 \otimes V_2) \oplus (V_2 \otimes V_1) \right] 
  = \text{span} \left\{ \uvec^1_i \otimes \uvec^2_j - \uvec^2_j \otimes \uvec^1_i \right\}_{i = 1, \dots N_\rmo, \, j = 1, \dots, N_2}.
\]
Following directly from the properties of the tensor product, we deduce that antisymmetric tensor product is symmetric, i.e $V_1 \wedge V_2 \cong V_2 \wedge V_1$, and distributive over direct sum, i.e, $V \wedge (V_1 \oplus V_2) = (V \wedge V_1) \oplus (V \wedge V_2)$. 

The fermionic many body Hilbert space can then be defined as a sum over sectors with fixed particle numbers:
\beq 
  \hilbert = \bigoplus_{N=0}^\infty \hilbert_{N}, \qquad 
  \hilbert_{N} \equiv \bigwedge_{n=1}^N \hilbert_1,   \label{eq:hilbert_def}
\eeq 
where $\hilbert_0 \cong\cmplx$ is spanned by the unique vacuum state, while $\hilbert_1$ is the single-particle Hilbert space. We seek to show that a subspace of $\hilbert_1$ naturally induces a tensor decomposition of $\hilbert$. To this end, consider a subspace $\hilbert_{1,\sA} \subset \hilbert_1$ with its orthogonal complement $\hilbert_{1,B}$, so that $\hilbert_1 = \hilbert_{1,\sA} \oplus \hilbert_{1,\sB}$. Substituting in Equation \ref{eq:hilbert_def} and using the symmetry and distributivity of $\wedge$, we get 
\begin{eqnarray}
 \hilbert & = \hilbert_0 \oplus \left[ \bigoplus_{N=1}^\infty  \bigwedge_{n=1}^N \left( \hilbert_{1,\sA} \oplus \hilbert_{1,\sB} \right) \right] \nonumber \\ 
 & = \hilbert_0 \oplus  \left[ \bigoplus_{N=1}^\infty \bigoplus_{m=0}^N  \left( \bigwedge_{n=0}^m \hilbert_{1,\sA} \right) \wedge \left( \bigwedge_{n=m+1}^N \hilbert_{1,\sB} \right)  \right], 
\end{eqnarray}
which is an infinite series with terms of the form $\hilbert_{1,\sA}^m \wedge \hilbert_{1,\sB}^n$ with nonnegative integers $m,n$, excluding $m=n=0$. Defining $\hilbert_{0, \sA/\sB} \cong \cmplx$, we set $\hilbert_0 = \hilbert_{0,\sA} \wedge \hilbert_{0,\sB}$, so that 
\beq
 \hilbert =  \left[ \hilbert_{0, \sA} \oplus \left( \bigoplus_{N_\sA=1}^\infty \bigwedge_{n=0}^{N_\sA} \hilbert_{1,\sA} \right) \right] \wedge \left[ \hilbert_{0, \sB} \oplus \left( \bigoplus_{N_\sB=1}^{\infty}  \bigwedge_{n=0}^{N_\sB} \hilbert_{1,\sB} \right) \right].
\eeq
Defining the many-body states corresponding to the single-particle Hilbert spaces $\hilbert_{0, \sA/\sB}$ analogous to Equation \ref{eq:hilbert_def}, we get 
\beq
 \hilbert =  \left( \bigoplus_{m=0}^\infty \hilbert_{m,\sA} \right) \wedge \left( \bigoplus_{n=0}^{\infty}  \hilbert_{n,\sB} \right)  = \hilbert_\sA \wedge \hilbert_\sB,
\eeq
which is the desired tensor decomposition of the fermionic many-body Hilbert space.

\section{Gramian matrices}   \label{app:gram}
Let $\vv_a \in \mathbb{V}, \, a = 1, \dots N$ be a set of vectors, where $\mathbb{V}$ is a complex vector space with a positive definite, sesquilinear inner product $\inner{}{}$. The \emph{Gramian} matrix $\overlap$ associated with these vectors is defined as 
\beq 
  \overlap_{ab} \equiv \inner{\vv_a}{\vv_b}.
\eeq 
Clearly, $\overlap$ is a Hermitian matrix, so that the spectrum $\spec{\overlap} \subset \real$, and the eigenbasis of $\overlap$ is orthonormal and spans $\cmplx^N$. Let $\bga\in\cmplx^N$, and consider 
\beq 
  \fl \qquad \bga^\dagger \overlap \bga 
  = \sum_{a,b=1}^N \gamma_a^\ast \inner{\vv_a}{\vv_b} \gamma_b^\pdg 
  = \inner{\sum_{a=1}^N \gamma_a \vv_a}{\sum_{b=1}^N \gamma_b \vv_b} 
  = \inner{\vv_\bga}{\vv_\bga} \geq 0,
\eeq 
since the inner product is positive-definite. Furthermore, the equality above holds only if $\vv_\gamma = 0$, i.e, if there exist a set of $\gamma_a \in \cmplx$ such that $\sum_a \gamma_a \vv_a = \nullv$, i.e, if the vectors are linearly dependent. Thus, we deduce that $\overlap \geq 0$, with the equality satisfied \emph{iff} the vectors $\vv_a$ are linearly independent. This provides a lower bound on $\spec{\overlap}$. In general, we can only deduce a trivial upper bound on the spectrum, \viz, $\overlap \leq \tr{\overlap}$. 

A better upper bound for the spectrum is possible in one practical case, \viz, when the vectors are obtained by projecting down an orthonormal set of vectors from some bigger Hilbert space. To wit, consider an orthonormal set of vectors $\mathbf{V}_a \in \mathbb{V}_0$, so that the associated overlap matrix is simply the $N\times N$ identity matrix. Let $\mathbb{V} \subset \mathbb{V}_0$ with the projector $\proj \cn \mathbb{V}_0 \to \mathbb{V}$, and define $\vv_a = \proj \mathbf{V}_a$, which are not necessarily orthonormal. However, the Hilbert space splits as $\mathbb{V}_0 = \mathbb{V} \oplus \mathbb{V}^\perp$, with the orthogonal projector $\proj^\perp \cn \mathbb{V}_0 \to \mathbb{V}^\perp$ defined as $\proj^\perp = \id - \proj$. Thus, we can define vectors $\vv_a^\perp = \proj^\perp \mathbf{V}_a$, and hence the two Grammian matrices 
\beq 
  \overlap_{ab} \equiv \inner{\vv_a}{\vv_b},  \qquad 
  \overlap^\perp_{ab} \equiv \inner{\vv^\perp_a}{\vv^\perp_b} = \left(\id - \overlap\right)_{ab}.
\eeq
Thus, we conclude that $\overlap \geq 0$ and $\id - \overlap \geq 0$, from which we can deduce that $\spec{\overlap} \subset [0,1]$.

\section{Hamiltonian vector fields and $\mathfrak{isp}(2d, \real)$}   \label{app:isp}
In the symplectic formulation of classical mechanics\cite{marsden2013introduction, arnold_book}, the time evolution of the classical system corresponds to the flows generated by Hamiltonian vector fields. 
Given a Hamiltonian $H \cn \real^{2d} \to \real$, one uses $\omega$ to define a Hamiltonian vector field $\mathfrak{X}_H = \mathfrak{X}_H^i \frac{\partial}{\partial \xi^i}$ on the phase space as 
\beq 
    dH = \omega(-, \mathfrak{X}_H) \iff \mathfrak{X}_H^i = (\bom^{-1})^{ij} \frac{\partial H}{\partial \xi^j},    \label{eq:symp_hlt}
\eeq 
Conversely, given a Hamiltonian vector field, i.e, one that preserves the symplectic form, one can use this relation to derive the corresponding Hamiltonian. More formally, for a symplectic manifold, the Lie algebra of tangent vectors is homomorphic to the Poisson algebra of Hamiltonians.

Such a set of Hamiltonian vector fields are defined by the action of the one-parameter families of symplectomorphisms on the phase space. In particular, for the phase space $\real^{2d}$, the group of linear symplectomorphisms is the Lie group $\ISp(2d, \real)$, whose Lie algebra $\mathfrak{isp}(2d, \real)$ is simply the tangent space of the group manifold at identity, alongwith the Lie algebra of tangent vectors. Thus, we can associate Hamiltonians to the elements of the Lie algebra. Explicitly, consider the one-parameter family of curves generated by $X \in \mathfrak{isp}(2d, \real)$, i.e, $\bxi(t) = e^{tX} \circ \bxi$, so that
\beq 
   \mathfrak{X}^j = \frac{d}{dt} \bxi^j(t) \big|_{t=0} = \left[X \circ \bxi\right]^j = \left[ \fA \bxi + \fb \right]^j
\eeq
Next, since $\bom^T = -\bom$ and $\fA \in \mathfrak{sp}(2d, \real)$, so that 
\beq    
  (\id + \epsilon \fA)^T \bom (\id + \epsilon \fA) = \bom 
  \implies \bom \fA = (\bom \fA)^T \quad \text{up} \; \text{to} \; \; O(\epsilon).
\eeq 
Thus, using Equation \ref{eq:symp_hlt} and the fact that  $\bom\fA$ is a symmetric matrix, we get 
\beq 
   \frac{\partial H}{\partial \xi^j} =  \left(\bom \fA \right)^{jk} \xi^k + \bom^{jk} b^k \implies 
   H = \frac{1}{2} \bxi^T  \bom \fA \, \bxi + \fb^T \bom \, \bxi,
\eeq
which is the closed form of the Hamiltonian that generates the same flow as the one-parameter subgroup $g(t) = \exp(tX_K)$.

\section{Wigner functions}    \label{app:wigner}
For a pure density matrix $\hat\rho = \ket{\psi}\bra{\psi}$, the density matrix is defined as 
\beq 
  \fl \qquad
  W(\vx, \vp) = \bra{\psi} \wigE(\vx,\vp) \ket{\psi} = \int \frac{ \rmd^dp_1 \rmd^dx_1 }{(2\pi)^d} \rme^{i(\vx \cdot \vp_1 - \vx_1 \cdot \vp)} \bra{\psi} \wige(\vx_1, \vp_1) \ket{\psi}.  
  \label{eq:wigner_def1}
\eeq
We seek to perform the computations in the position basis $\{ \ket{x}, x \in\real^d \}$, with orthogonality and completeness relations 
\beq 
   \braket{\vx}{\vx'} = \delta^d (\vx - \vx'),
   \qquad \id = \int_{\real} \rmd^d x \, \ket{\vx} \bra{\vx}.
\eeq 
Inserting this resolution of identity in \eq{eq:wigner_def1} and using $\psi(\vx) = \braket{\vx}{\psi}$, we get 
\[
 \fl \qquad  W(\vx, \vp) = \frac{1}{(2\pi)^d} \int \rmd^dp_1 \rmd^dx_1 \rmd^dx_2 \rmd^dx_3 \,  \rme^{i(\vx \cdot \vp_1 - \vx_1 \cdot \vp)} \psi^\ast(\vx_2) \psi(\vx_3) \bra{\vx_2} \wige(\vx_1, \vp_1) \ket{\vx_3}
\]
Using the defintion of $\wige(\vx, \vp)$ from Equation \ref{eq:def_wige}, we get 
\[
 \fl \qquad \bra{\vx_2} \wige(x_1, p_1) \ket{\vx_3} 
 = \rme^{- \frac{i}{2} \vx_1 \cdot \vp_1 } \bra{\vx_2} \rme^{ - i \vp_1 \cdot \hat{\vx}} \rme^{i \vx_1 \cdot \hat{\vp}} \ket{\vx_3} 
 =  \rme^{-i \vp_1 \cdot \left( \frac{1}{2} \vx_1  + \vx_2 \right) } \delta^d(\vx_1 + \vx_2 - \vx_3). 
\]
The Wigner function becomes 
\begin{eqnarray}
 \fl \qquad W(\vx, \vp) & = \frac{1}{(2\pi)^d} \int \rmd^dp_1 \rmd^dx_1 \rmd^dx_2 \rmd^dx_3 \, \psi^\ast(\vx_2) \psi(\vx_3)  \rme^{i \vp_1 \cdot \left( \vx - \frac{1}{2} \vx_1  - \vx_2 \right) - i \vp \cdot \vx_1}  \delta^d(\vx_1 + \vx_2 - \vx_3).  \nonumber \\
 \fl & = \int \rmd^dx_1 \rmd^dx_2 \, \rme^{- i \vp \cdot \vx_1} \psi^\ast(\vx_2) \psi(\vx_1 + \vx_2) \int \frac{\rmd^dp_1}{(2\pi)^d} \rme^{i \vp_1 \cdot \left( \vx - \frac{1}{2} \vx_1  - \vx_2 \right) }  \nonumber \\
 \fl & = \int \rmd^dx_1 \, \rme^{- i \vp \cdot \vx_1} \psi^\ast\left( \vx - \frac{\vx_1}{2} \right) \psi\left( \vx + \frac{\vx_1}{2} \right).
\end{eqnarray}
For the inverse transformation, we again insert resolutions of identity, and compute
\begin{eqnarray}
  \bra{\vx_2} \wigE(\vx,\vp) \ket{\vx_3}
  & = \int \frac{\rmd^dx_1 \rmd^dp_1}{(2\pi)^d}  \rme^{i \vp_1 \cdot \left( \vx - \frac{1}{2} \vx_1  - \vx_2 \right) - i \vp \cdot \vx_1}  \delta^d(\vx_1 + \vx_2 - \vx_3).  \nonumber \\
  & = \rme^{i \vp \cdot (\vx_2 - \vx_3)} \int \frac{\rmd^dp_1}{(2\pi)^d} \rme^{ - i \vp_1 \cdot \left( \vx - \frac{1}{2} (\vx_2 + \vx_3) \right) } \nonumber \\
  & = \rme^{i \vp \cdot (\vx_2 - \vx_3)} \delta^d \left( \vx - \frac{\vx_2 + \vx_3}{2} \right)  \label{eq:expt_cont}.
\end{eqnarray}
Hence, 
\begin{eqnarray}
 \hat\rho & =  \int \frac{\rmd^dx \, \rmd^dp}{(2\pi)^d} W(\vx, \vp) \int \rmd^dx_2 \rmd^dx_3 \, \ket{\vx_2} \bra{\vx_2} \wigE(\vx, \vp) \ket{\vx_3} \bra{\vx_3}  \nonumber \\
 & =  \int \rmd^dx_2 \, \rmd^dx_3  \left[ \int \frac{\rmd^dp}{(2\pi)^d}  \,  W\left(\frac{\vx_2 + \vx_3}{2}, \vp\right) \,  \rme^{i \vp \cdot (\vx_2 - \vx_3)} \right] \ket{\vx_2}  \bra{\vx_3},  
\end{eqnarray}
from which the new wavefunctions can be read off using the definition of the density matrix operator.

\section{The Fractional Fourier Transform}  \label{app:fracFT}
The phase space rotation corresponds to the convolution of a wavefunction by the kernel
\beq
  \uop_\theta(x,y) = \frac{\rme^{i\phi(\theta)}}{\sqrt{2\pi|\sin\theta|}} \rme^{ - \frac{i}{2} \cot\theta (x^2  - 2x y \sec\theta + y^2) }, 
\eeq
as shown in Section \ref{sec:rot_fracFT}. A particularly nice choice of $\phi(\theta)$ is to demand that composition of kernels adds the corresponding $\theta$, i.e, 
\beq 
  \int_\real \rmd z \, \uop_\theta(x,z) K_{\theta'}(z,y) = K_{\theta+\theta'}(x,y). 
\eeq
The LHS can be evaluated as 
\begin{eqnarray}
  \fl  & \qquad \frac{\rme^{i\phi(\theta) + i\phi(\theta')}}{2\pi\sqrt{|\sin\theta \sin\theta'|}} 
  \int_\real \rmd z \, \rme^{ - \frac{i}{2} \cot\theta (x^2  - 2x z \sec\theta + z^2) - \frac{i}{2} \cot\theta' (z^2  - 2y z \sec\theta + y^2) } \nonumber \\ 
  \fl  & = \frac{\rme^{i \left[\phi(\theta) + \phi(\theta') \right]}}{2\pi\sqrt{|\sin\theta \sin\theta'|}} \rme^{ - \frac{i}{2} \left( x^2 \cot\theta + y^2 \cot\theta' \right)}  
  \int_\real \rmd z \, \rme^{ - \frac{i}{2} (\cot\theta + \cot\theta') z^2 + i (x \csc\theta + y \csc\theta') z } \nonumber \\ 
  \fl  & = \frac{\rme^{i \left[\phi(\theta) + \phi(\theta') \right]}}{2\pi\sqrt{|\sin\theta \sin\theta'|}} \rme^{ - \frac{i}{2} \left( x^2 \cot\theta + y^2 \cot\theta' - \frac{(x \csc\theta + y \csc\theta')^2}{\cot\theta+\cot\theta'}\right)}  
  \int_\real \rmd z \, \rme^{ - \frac{i}{2} (\cot\theta + \cot\theta') \left(z - 	 \frac{x \csc\theta + y \csc\theta'}{\cot\theta+\cot\theta'} \right)^2}  \nonumber \\ 
  \fl  & = \frac{\rme^{i \left[\phi(\theta) + \phi(\theta') \right]}}{\pi\sqrt{|2\sin\theta \sin\theta'(\cot\theta+\cot\theta')|}} \rme^{ - \frac{i}{2} \left( x^2 \cot\theta + y^2 \cot\theta' - \frac{(x \csc\theta + y \csc\theta')^2}{\cot\theta+\cot\theta'}\right)} \int_\real du \, \rme^{ - iu^2} \nonumber \\ 
  \fl  & = \frac{\rme^{i \left[\phi(\theta) + \phi(\theta') \right]}}{\pi\sqrt{|2\sin(\theta+\theta')|}} \rme^{ - \frac{i}{2} \left[ \left( x^2 + y^2 \right) \frac{\cot\theta\cot\theta' -1}{\cot\theta+\cot\theta'}  - 2xy \frac{\csc\theta\csc\theta'}{\cot\theta+\cot\theta'} \right] } \, \rme^{-i\pi/4} \sqrt{\pi}  \nonumber \\ 
  \fl  & = \frac{\rme^{i \left[\phi(\theta) + \phi(\theta') - \pi/4 \right]}}{\sqrt{|2\pi\sin(\theta+\theta')|}}  \rme^{ - \frac{i}{2} \cot(\theta + \theta') \left( x^2  - 2x y \sec(\theta + \theta') + y^2 \right) }
\end{eqnarray}
where we have used 
\[
  \sin\theta \sin\theta'(\cot\theta+\cot\theta') = \cos\theta\sin\theta' + \sin\theta\cos\theta' = \sin(\theta + \theta')
\]
and the integral
\beq
  \int_\real du \, \rme^{ - iu^2} =  \rme^{-i\pi/4} \int_0^\infty dv \, \rme^{-v^2} = \rme^{-i\pi/4} \sqrt{\pi}  \label{eq:cont_int}
\eeq
with $v = \rme^{i\pi/4}u$ by a rotation of the integration contour by $\pi/4$ clockwise in the complex-$u$ plane. Thus, we must demand that 
\beq 
  \phi(\theta + \theta') = \phi(\theta) + \phi(\theta') - \frac{\pi}{4},
\eeq 
so that $\phi(\theta)-\frac{\pi}{4}$ must be linear in $\theta$. Finally, demanding that $\uop_\theta(x,y)$ reproduce the Fourier transform kernel, i.e, $K_{\pi/2}(x,y) = \rme^{ixy}$, to get $\phi\left( \frac{\pi}{2} \right) = 0$. We set $\phi(\theta) = \frac{\pi}{4} - \frac{\theta}{2}$, which reduces $\uop_\theta$ to the kernel for the \emph{fractional Fourier transform}. It can alternatively be written as 
\beq 
  \uop_\theta (x,y) = \sqrt{ \frac{1-i\cot\theta}{2\pi} }  \rme^{ - \frac{i}{2} \cot\theta (x^2  - 2x y \sec\theta + y^2) }.
\eeq
We can derive the following limiting cases: 
\beq  
  \uop_\theta(x,y) = \cases{
    \delta\left(x - \zeta_n y\right) & $\theta = n\pi$\\
    \frac{1}{\sqrt{2\pi}} \, \rme^{i \zeta_n xy} & $\theta = (2n+1) \frac{\pi}{2}$\\},
\eeq
where $\zeta_n \equiv \cos(n\pi) = (-1)^n$. The latter case directly follows from setting $\theta = (2n+1) \frac{\pi}{2}$, while for the former, we need to take the limit: 
\beq 
    K_{n\pi}(x,y) = \lim_{\epsilon\to 0} K_{n\pi + \epsilon} (x,y) = \frac{\rme^{i\pi/4}}{\sqrt{\pi}}  \lim_{\epsilon \to 0} \frac{1}{\sqrt{2\epsilon}} \, \rme^{ - \frac{i}{2\epsilon} \left( x - \zeta_n y \right)^2 }.
\eeq 
This limit is defined as a distribution, which can be evaluated as 
\begin{eqnarray}
  \fl \qquad \int_{-\infty}^\infty \rmd y \, K_{n\pi}(x,y) f(y) 
  & = \frac{\rme^{i\pi/4}}{\sqrt{\pi}}  \lim_{\epsilon \to 0} \int_{-\infty}^\infty \frac{\rmd y}{\sqrt{2\epsilon}} \, \rme^{ - \frac{i}{2\epsilon} \left( x - \zeta_n y \right)^2 } f(y) \nonumber \\ 
  \fl \qquad 
  & = \frac{\rme^{i\pi/4}}{\sqrt{\pi}}  \lim_{\epsilon \to 0} \int_{-\infty}^\infty \rmd u \, \rme^{ - i u^2} f\left( \epsilon u + \zeta_n x \right) \nonumber \\ .
  \fl \qquad 
  & = f\left( \zeta_n x \right) \frac{\rme^{i\pi/4}}{\sqrt{\pi}}  \int_{-\infty}^\infty \rmd u \, \rme^{ - i u^2} = f\left( \zeta_n x \right),
\end{eqnarray}
where $u \equiv \left( y - \zeta_n x \right)/\sqrt{2 \epsilon}$ and we have used Equation \ref{eq:cont_int} in the last step.

\section{Harmonic oscillator Wigner functions}     \label{app:HO_wigner}
We seek to compute
\beq 
  \fl \qquad W_{mn}(\coord) = \bra{\varphi_m} \wigE(\coord) \ket{\varphi_n} 
  = \frac{1}{2\pi} \int \rmd^2\coord_1 \; \rme^{\coord^\ast \coord_1 - \coord_1^\ast \coord} \;\rme^{-|\coord_1|^2/2} \bra{\varphi_m} \rme^{-\coord_1 \ad} \rme^{\coord_1^\ast \a} \ket{\varphi_n},
\eeq 
where $\rmd^2\coord_1 = i \rmd\coord_1 \rmd\coord_1^\ast$. The energy eigenstates of the harmonic oscillator, $\ket{\varphi_n}$ satisfy:
\beq 
 \a \ket{\varphi_n} = \sqrt{n} \; \ket{\varphi_{n-1}} \qquad
 \ad \ket{\varphi_n} = \sqrt{n+1} \; \ket{\varphi_{n+1}}.
\eeq
Thus, the expectation value can be evaluated to get 
\beq 
 \bra{\varphi_m} \rme^{-\coord_1 \ad} \rme^{\coord_1^\ast \a} \ket{\varphi_n} =  \sqrt{m! n!} \sum_{q=0}^{\text{min}(m,n)} \frac{(-\coord_1)^{m-q} (\coord_1^\ast)^{n-q}}{q! (m-q)! (n-q)!},
\eeq
where we have used 
\[
 \rme^{\coord_1^\ast \a} \ket{\varphi_n} = \sum_{k=0}^\infty \frac{(\coord_1^\ast)^k}{k!} \a^k \ket{\varphi_n} = \sum_{q=0}^n \frac{(\coord_1^\ast)^{n-q}}{(n-q)!} \sqrt{\frac{n!}{q!}} \, \ket{\varphi_q}.
\]
Setting $\coord = \frac{r}{\sqrt{2}} \rme^{i\theta}$ and $\nu = n-m$, this simplifies to 
\beq 
  \fl \qquad \bra{\varphi_m} \rme^{-\coord_1 \ad} \rme^{\coord_1^\ast \a} \ket{\varphi_n} = \rme^{i\nu\theta_1} \sqrt{m! n!}  \sum_{q=0}^{\text{min}(m,n)} \frac{ (-1)^{m-q}}{q! (m-q)! (n-q)!} \left( \frac{r_1}{\sqrt{2}} \right)^{m+n-2q}.
\eeq 
These sums evaluate to the associated Laguerre polynomials\cite{bateman_int2}, so that 
\beq 
  \bra{\varphi_m} \rme^{-\coord_1 \ad} \rme^{\coord_1^\ast \a} \ket{\varphi_n} = \rme^{i\nu\theta_1} \sqrt{\frac{m!}{n!}} \left( \frac{r_1}{\sqrt{2}} \right)^\nu L^{\nu}_{m} \left(\frac{r_1^2}{2}\right)
\eeq 
Thus, 
\beq 
  \fl \qquad W_{mn} (r, \theta) = \sqrt{\frac{m!}{n!}}  \int \frac{r_1 \rmd r_1 \rmd\theta_1}{2\pi} \rme^{-\frac{r_1^2}{2} + i (\nu\theta_1 + r r_1 \sin(\theta_1 - \theta))}  \left( \frac{r_1}{\sqrt{2}} \right)^\nu L^{\nu}_{m} \left(\frac{r_1^2}{2}\right).
\eeq
The $\theta_1$ integral can be evaluated using the definition of the Bessel function of the first kind: 
\[ 
  J_\nu(a) = \int_0^{2\pi} \frac{\rmd\phi}{2\pi} \rme^{i \left( \nu\phi + a\sin\phi \right)}, 
\]
so that 
\[
  \fl \qquad W_{mn}(r, \theta) = \rme^{i \nu \theta} \sqrt{\frac{m!}{n!}} \int_0^\infty  \rmd r_1 \sqrt{r r_1} f(r_1) J_\nu (r r_1); \quad 
  f(r_1) =  \frac{r_1^{\nu + \frac{1}{2}}  \rme^{-\frac{r_1^2}{4}}}{ 2^{\frac{\nu}{2}}}  L^{\nu}_{m} \left( \frac{r_1^2}{2} \right).
\]
We identify this as a Hankel transform (Bateman manuscript, vol II, page 43, eqn 5)\cite{bateman_int2}, so that 
\[
 W_{mn}(r,\theta) = 2 (-1)^m  \sqrt{\frac{m!}{n!}}   \left( \sqrt{2} \, r \rme^{i \theta} \right)^\nu \rme^{-r^2} L^\nu_m (2 r^2) \label{eq:Wmn_r}.
\]
Switching back to complex coordinates $\coord = \frac{1}{\sqrt{2}} \left( x + i p \right)$, we get
\beq  
   W_{mn} (\coord) = 2 (-1)^m \sqrt{\frac{m!}{n!}} (2\coord)^{n-m} \rme^{-2 |\coord|^2} L^{n-m}_m (4|\coord|^2),
\eeq
which is the desired result.

To compute the overlap on half space between the harmonic oscillator wavefunctions, we seek to compute 
\beq 
  \fl \qquad \overlapHO_{mn} =  \int_0^\infty \rmd x \int_{-\infty}^\infty \frac{\rmd p}{2\pi} W_{mn} (x, p) =  \int_0^\infty r \rmd r \int_{-\pi/2}^{\pi/2} \frac{\rmd\theta}{2\pi} W_{mn} (r, \theta), 
\eeq
where the polar representation is particularly suited for the case in hand. The angular integral evaluates to $\frac{1}{2}$ for $\nu = 0$, and to
\beq 
   I_\theta \equiv \int_{-\pi/2}^{\pi/2} \frac{\rmd\theta}{2\pi} \rme^{i\nu\theta} = \frac{1}{\pi\nu} \sin\left( \frac{\nu\pi}{2} \right) = \cases{ 
   	\frac{(-1)^{\frac{\nu-1}{2}}}{\pi \nu}, & $\nu \in 2\intg + 1$, \\ 
   	0, & $\nu \in 2\intg$, } 
\eeq
for $\nu \neq 0$. This expression does tend to $\frac{1}{2}$ as $\nu\to 0$. For the radial part, we need to evaluate
\begin{eqnarray*}
 \fl \qquad \int_0^\infty 2r \rmd r \; r^\nu \rme^{-r^2}  L^\nu_m (2 r^2) 
 & = \int_0^\infty \rmd u \; u^{\nu/2} \rme^{-u}  L^\nu_m (2 u) \\ 
 \fl \qquad  & = \sum_{q=0}^m  \frac{(-2)^q}{q!} \binom{m+\nu}{q + \nu} \int_0^\infty du \; u^{\nu/2 + q} \rme^{-u}  \\ 
 \fl \qquad  & = \binom{m+\nu}{m} \, \Gamma\left( \frac{\nu}{2} + 1 \right)  \hypgeo \left( -m, \frac{\nu}{2} + 1, \nu + 1; 2 \right) 
\end{eqnarray*} 
where $u = r^2$, and we have used the definition of the Laguerre polynomials. Here, $\hypgeo$ denotes the ordinary hypergeometric functions, defined as 
\beq 
  \hypgeo(a,b,c; x) = \sum_{q=0}^\infty \frac{(a)_q (b)_q}{(c)_q} \frac{x^q}{q!},
\eeq 
where the the Pochhammer symbol is
\beq 
  (n)_q \equiv \cases{  1, & $q = 0$, \\ n(n+1) \dots (n + q - 1), & $q > 0$. }.
\eeq 
Putting all the pieces together, the half space integral becomes 
\beq 
  \fl \qquad \overlapHO_{mn} = \sqrt{\frac{n!}{m!}}  \frac{(-1)^m 2^{\nu/2} \Gamma\left( \frac{\nu}{2} \right) }{\pi \Gamma(\nu+1)}  \hypgeo \left( -m, \frac{\nu}{2} + 1, \nu+1; 2 \right) \sin\left( \frac{\nu\pi}{2} \right) ,
\eeq
with $\nu = n-m$, and a $\lim_{\nu \to 0}$ is understood since this expression is undefined for $\nu = 0$. Although it is not immediately obvious, the expression is indeed symmetric under $m \leftrightarrow n, \, \nu \to -\nu$.

\section{Coherent states Wigner functions} 
\newcommand{\wcoord}{w}
A particular case, where the phase space rotations are analytically tractable, as well as easily visualizable, are the harmonic oscillator coherent states. We present that computation here. 

\begin{figure}[ht]
  \centering
  \includegraphics[width=0.3\textwidth]{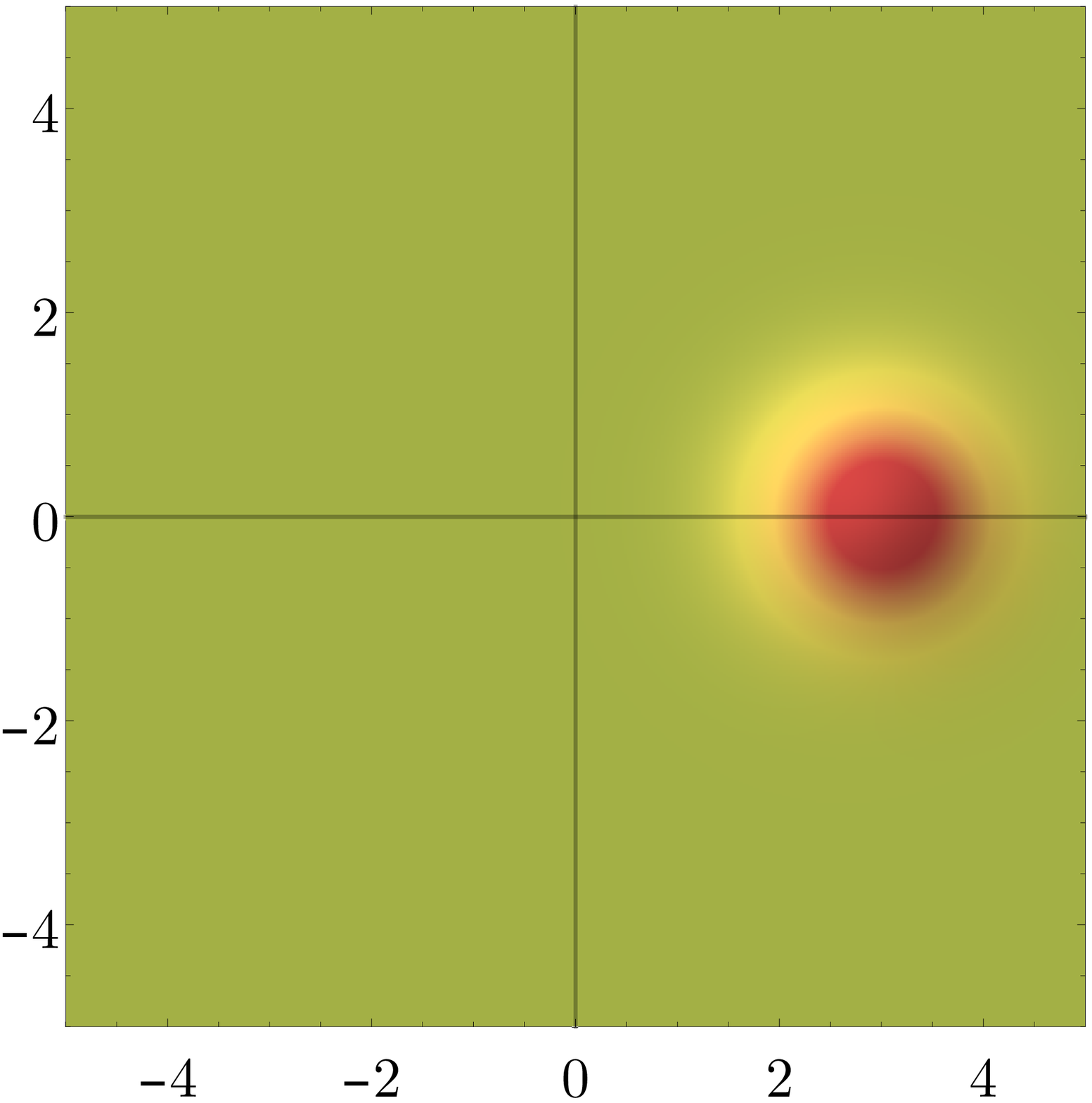} 
  \includegraphics[width=0.3\textwidth]{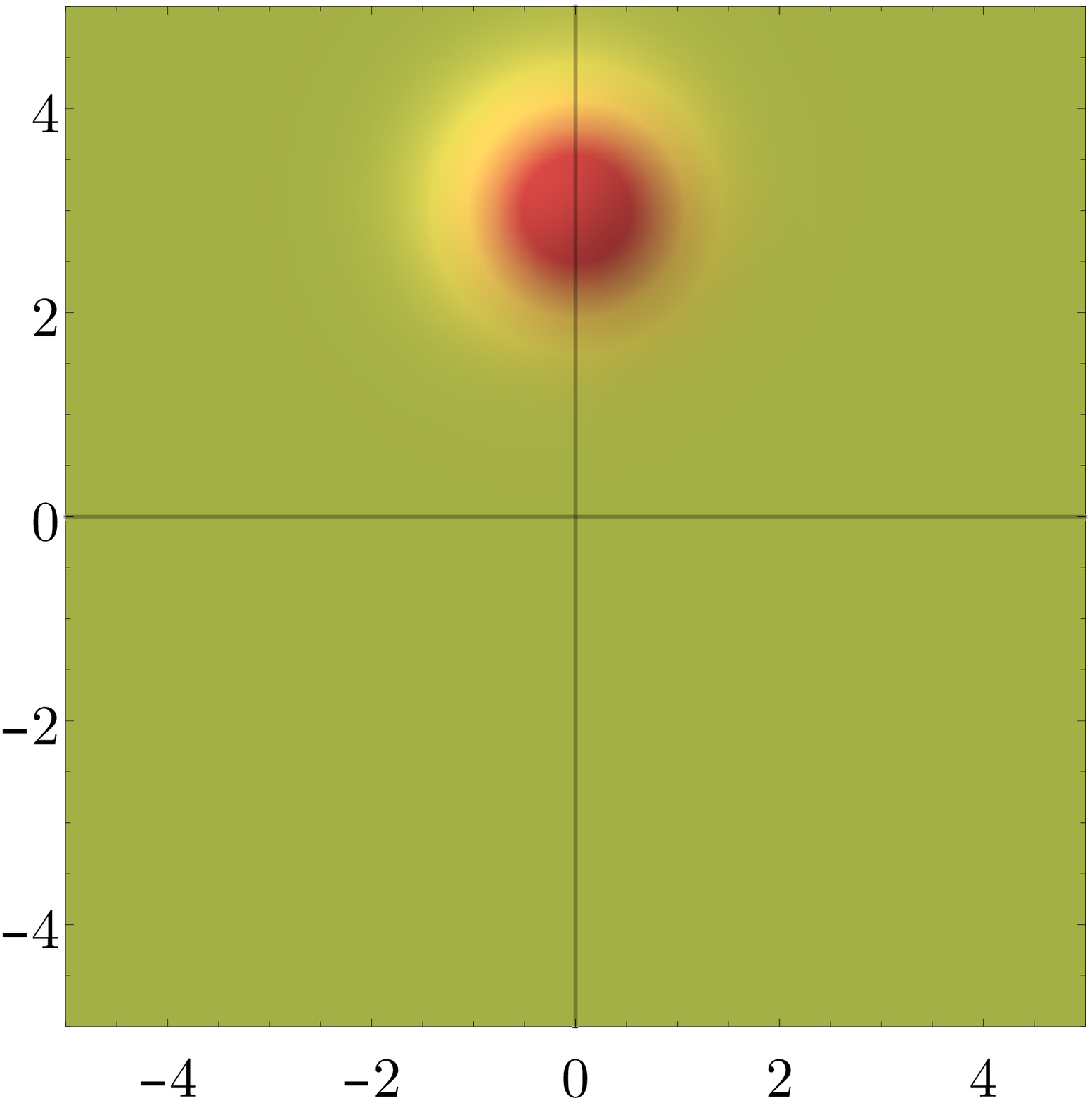}  
  \includegraphics[width=0.06\textwidth]{legend.eps} 
  \caption{The Wigner functions for the coherent states with $w = 3$ and $w = 3i$, respectively. } 
  \label{fig:wignerCo}
\end{figure}

The coherent states are the eigenstates of the harmonic oscillator annihilation operator ($\a \ket{\wcoord} = \wcoord \ket{\wcoord}$), more explicitly defined as 
\beq 
   \ket{\wcoord} = \rme^{\wcoord \ad - \wcoord^\ast \a} \ket{0} = \rme^{-\frac{1}{2} \abs{\wcoord}^2} \rme^{\wcoord \ad} \ket{0}. 
\eeq
where $\ket{0}$ is the vacuum state for the harmonic oscillator, i.e, $\a\ket{0} = 0$. The Wigner function is 
\begin{eqnarray}
  W_\wcoord(\coord) 
  & = \frac{1}{2\pi} \int \rmd^2 \coord_1 \; \rme^{\coord^\ast \coord_1 - \coord_1^\ast \coord - \frac{1}{2} \abs{\coord_1}^2 - \abs{\wcoord}^2} \bra{0} \rme^{\wcoord^\ast \a} \rme^{-\coord_1 \ad} \rme^{\coord_1^\ast \a} \rme^{\wcoord \ad} \ket{0},  \nonumber \\ 
  & = \frac{1}{2\pi} \int \rmd^2 \coord_1 \; \rme^{(\coord^\ast -\wcoord^\ast) \coord_1 - \coord_1^\ast (\coord-\wcoord) - \frac{1}{2} \abs{\coord_1}^2} = 2 \rme^{-2\abs{\coord-\wcoord}^2},
\end{eqnarray}
i.e, a Gaussian centered at $\wcoord$. Under phase space rotations, 
\[
  \ket{\wcoord} \to \rme^{i \theta \ad\a} \ket{\wcoord} = \rme^{-\frac{1}{2} \abs{\wcoord \rme^{i\theta}}^2} \sum_{n=0}^\infty \frac{\left( \wcoord \rme^{i \theta} \right)^n}{n!} \left( \ad \right)^n \ket{0} = \ket{\wcoord \rme^{i \theta}}.
\]
Thus, the coherent state rotates around the origin at a fixed radius $\abs{\wcoord}$. This can also be seen by a glance at the Wigner function, as shown in Fig \ref{fig:wignerCo}.

\section*{References}

\bibliographystyle{iopart-num}
\bibliography{ent_phsp}

\end{document}